\RequirePackage[l2tabu, orthodox]{nag}
\RequirePackage{snapshot}

\documentclass[9pt,onecolumn]{extarticle}

\makeatletter
\if@twocolumn
  \usepackage[dvips,letterpaper,top=0.5in, bottom=0.5in, left=0.75in, right=0.5in,includefoot,heightrounded]{geometry}
\else
  \usepackage[dvips,letterpaper,margin=1in,includefoot,heightrounded]{geometry}
\fi

\usepackage{srcltx}

\usepackage[russian,portuges,english]{babel}

\iflanguage{portuges}
    {\newcommand{\keywordname}{Palavras-chaves}}
    {\newcommand{\keywordname}{Keywords}}

\usepackage{amsmath}
\usepackage{amssymb,amsfonts}

\usepackage{abstract}

\usepackage{graphicx}
\usepackage[usenames,dvipsnames,svgnames,x11names]{xcolor}
\usepackage{subfigure}

\usepackage[squaren]{SIunits}

\usepackage{booktabs}

\usepackage{setspace}
\usepackage{flushend}

\usepackage{cite}

\usepackage{hyperref}
\usepackage[normalem]{ulem}

\usepackage{enumerate}

\usepackage{multirow}
\usepackage[noend]{algpseudocode}

\usepackage{listings}

\usepackage[short,12hr]{datetime}
       \usepackage{fouriernc}
\makeatletter

\newcommand{\printtitle}{%
\makeatletter
\if@twocolumn

\twocolumn[%
  \maketitle
  \begin{onecolabstract}
    \myabstract
  \end{onecolabstract}
  \begin{center}
    \small
    \textbf{\keywordname}
    \\\medskip
    \mykeywords
  \end{center}
  \bigskip
]
\saythanks
\else
  \maketitle
  \begin{onecolabstract}
    \myabstract
  \end{onecolabstract}
  \begin{center}
    \small
    \textbf{\keywordname}
    \\\medskip
    \mykeywords
  \end{center}
  \bigskip
  \onehalfspacing
\fi
\makeatother
}

\author{
T. L. T. da Silveira%
\thanks{Centro de Ci\^encias Computacionais, Universidade Federal do Rio Grande, Rio Grande, Brazil, e-mail: \texttt{tltsilveira@furg.br}}
\and
D. R. Canterle%
\thanks{Instituto de Matem\'atica e Estat\'{\i}stica, Universidade de S\~ao Paulo, S\~ao Paulo, Brazil.}
\thanks{Programa de P\'os-Gradua\c{c}\~ao em Engenharia El\'etrica, Universidade Federal de Pernambuco, Recife, Brazil, e-mail: \texttt{diegocanterle@gmail.com}}
\and
D.~F.~G.~Coelho%
\thanks{Independent researcher, Calgary, Canada, e-mail: \texttt{diegofgcoelho@gmail.com}}
\and
V.~A.~Coutinho%
\thanks{Departamento de Computa\c{c}\~ao, Universidade Federal Rural de Pernambuco, Recife, Brazil, e-mail: \texttt{vitor.coutinho@ufrpe.br}}
\and
F. M. Bayer%
\thanks{Departamento de Estatística and LACESM, Universidade Federal de Santa Maria, Santa Maria, Brazil, e-mail: \texttt{bayer@ufsm.br}}
\and
R.~J.~Cintra%
\thanks{Signal Processing Group,
Departamento de Estat\'{\i}stica, Universidade Federal de Pernambuco, Recife, Brazil, e-mail: \texttt{rjdsc@de.ufpe.br}}
}

\title{%
A Class of Low-Complexity DCT-like Transforms for Image and Video Coding}

\newcommand{\myabstract}{%
The discrete cosine transform (DCT) is a relevant tool in signal processing applications,
mainly known for
its good decorrelation properties.
Current image and video coding standards---such as JPEG and HEVC---adopt the DCT as a fundamental building block for compression.
Recent works have introduced low-complexity approximations for the DCT, which become paramount in applications demanding real-time computation and low-power consumption.
The design of DCT approximations involves
a trade-off between computational complexity and performance.
This paper introduces a new multiparametric transform class encompassing the round-off DCT (RDCT) and the modified RDCT (MRDCT), two relevant multiplierless 8-point approximate DCTs.
The associated fast algorithm is provided.
Four novel orthogonal low-complexity 8-point DCT approximations are obtained by solving a multicriteria optimization problem.
The optimal 8-point transforms are scaled to lengths 16 and 32
while keeping the arithmetic complexity low.
The proposed methods
are assessed
by
proximity and coding measures
with respect to the exact DCT.
Image and video coding experiments
hardware realization are performed.
The
novel
transforms perform close to or outperform the current state-of-the-art
DCT approximations.
}

\newcommand{\mykeywords}{%
DCT approximation,
low-complexity transforms,
image and video coding
}

\date{}

\begin{document}

\printtitle

\section{Introduction}

Discrete transforms play a central role in digital signal processing tasks, such as analysis, filtering, and coding~\cite{Oppenheim1999}.
Remarkable transforms include the discrete Hartley transform, the Haar transform, the discrete Fourier transform, the Karhunen-Lo\`eve transform (KLT), and the discrete sine and cosine transforms~\cite{V.Britanak2006, Strawderman1999, Ahmed1975}.
In the context of signal encoding, the KLT holds optimal decorrelation properties~\cite{V.Britanak2006} though being data-dependent~\cite{Tran2000} and, thus, computationally complex and expensive~\cite{Tran2000, Ahmed1974}.
Nonetheless, the discrete cosine transform (DCT), which is signal-independent,
performs close to optimal when applied to a high correlated first-order Markov random process~\cite{HsiehS.Hou1987,Lee1994}.
Natural images belong to this particular statistical class~\cite{Rao1990}, which makes efficient implementations of the DCT to be widely adopted in current image and video coding standards~\cite{ochoa2019discrete}.

Block transformation, fast algorithms, and approximate
computing
are some of the approaches for reducing the computational cost of the DCT applications~\cite{V.Britanak2006}.
Traditional image and video coding standards---such as the Joint Photography Experts Group (JPEG)~\cite{Wallace1992}
and the High Efficiency Video Coding (HEVC)~\cite{Sullivan2012}---operate in a block-based fashion, where the input signal is firstly segmented into disjoint blocks and then transformed accordingly~\cite{Wien2014}.
For instance, JPEG uses the 8-point DCT~\cite{Wallace1992}, whereas HEVC adopts transforms of length 4, 8, 16, and 32 for taking advantage of highly-correlated image parts of different sizes~\cite{MahsaT.Pourazard2012}.
Applying the DCT on an image block may result in few localized, meaningful coefficients that are further quantized. The low-frequency non-separable transform (LFNST)~\cite{koo2019} explicitly discards high-frequency coefficients before quantization for reducing memory usage and computation in the new Versatile Video Coding (VVC)~\cite{ZhaoTCSVT2021}.
Because of its relevance on image compression, several fast algorithms for the DCT calculation are reported in literature~\cite{Chen1977,Loeffler1989,Lee1994,HsiehS.Hou1987,Vetterli1984,Makhoul1980}. These approaches commonly explore sparse matrix factorizations~\cite{Chen1977,Loeffler1989}, recursiveness~\cite{Lee1994,HsiehS.Hou1987}, and
relationships with  other transforms~\cite{Vetterli1984,Makhoul1980}.
These efforts resulted in algorithms
that attain the theoretical
minimum
multiplicative
complexity~\cite{heideman1988multiplicative}, being nowadays a quite mature research area.

Even considering these algorithms,
the
DCT-based
transformation requires irrational quantities to be computed and stored, increasing the complexity of both encoders and decoders~\cite{ochoa2019discrete}.
Irrational quantities are often represented as floating-point in modern computers~\cite{itoh2017fast},
however,
such floating-point dependence may jeopardize the application of the DCT in very low-power scenarios~\cite{Harize2013,Ernawan2011}.
Internet of Things~(IoT) applications~\cite{alarifi2020novel,wu2018energy}
often employ low-power wide area networks (LPWANs) for video transmission.
Most IoT devices are designed at the lowest hardware cost, smaller hardware size, and the lowest battery consumption possible, and the LPWANs present narrow bandwidths.
Low-complexity data compression mechanisms become crucial in such cases.
In the past few years, several works proposed low-complexity DCT \emph{approximations}, mostly of length 8, capable of compromising between arithmetic complexity and coding efficiency~\cite{Cintra2011,Bayer2012a,Haweel2001,Bouguezel2011,Tablada2015,Bouguezel2008,SaadBouguezel2009,Bouguezel2013,Lengwehasatit2004,Bouguezel2010,Oliveira2018}.
Most of these linear transformation matrices have entries defined over the set $\mathcal{C} = \{0,\pm \frac{1}{2},\pm 1, \pm2\}$, and, thus, can be implemented using only addition and bit-shifting operations~\cite{Cintra2011, Bouguezel2011, Coutinho2016}.
Indeed, multiplication-free transforms are remarkable for reducing circuitry, chip area, and power consumption in hardware implementations~\cite{Potluri2014,Jridi2015}.

Prominent 8-point approximate DCTs include the
signed DCT
\cite{Haweel2001},
the series of transforms proposed by Bouguezel-Ahmad-Swamy
\cite{Bouguezel2011,Bouguezel2008,SaadBouguezel2009,Bouguezel2013,Bouguezel2010},
the Lengwehasatit-Ortega transform~\cite{Lengwehasatit2004},
the angle similarity-based DCT approximation~\cite{Oliveira2018},
the classes of transforms from~\cite{Tablada2015}~and~\cite{Canterle2020},
and, especially,
the round-off DCT (RDCT)~\cite{Cintra2011},
and the modified RDCT (MRDCT)~\cite{Bayer2012a}.
On the one hand, the RDCT outperforms the current state-of-the-art low-complexity  DCT approximations in terms of energy compaction properties~\cite{Jridi2015,Oliveira2017} at the expense of 22 additions.
On the other hand, the MRDCT is a \emph{very} low-complexity DCT-like transform that requires only 14 additions, the lowest arithmetic cost among the meaningful approximate DCTs archived in literature~\cite{Coutinho2016,DaSilveira2017a}.
Other very low-complexity approaches include the 14-additions transform from~\cite{Potluri2014} and the pruned MRDCT~\cite{Coutinho2016}, but they
perform poorer in image coding applications.

Overall, the current literature still lacks
unifying schemes that encompass known low-complexity DCT approximations.
Some authors, although, found that such formalizations might be useful for exploring structural properties and proposing novel, more powerful, transforms~\cite{Tablada2015,Canterle2020,Coelho2018}. Most of the classic related works focus on proposing 8-point DCT approximations only. Nowadays, however, there is a need for larger transforms for coping with high-resolution data~\cite{Sullivan2012,Jridi2015}. Some works proposed 16-point transforms~\cite{Bayer2012,DaSilveira2016a,DaSilveira2017a},  but none focused on larger blocklengths. To the best of our knowledge, there are only a few works exploring scalable approximate DCTs~\cite{Bouguezel2010,Bouguezel2013,Jridi2015,Oliveira2018,Canterle2020}.

Our contributions are the following.
First, we introduce a new class of low-complexity DCT-like transforms using a multiparametric formulation that encompasses the RDCT and the MRDCT as particular cases.
Our formalism explores the underlying search space aiming at transforms with both low-complexity and good energy compaction properties.
Second, we search for optimal transforms through a multicriteria optimization procedure and introduce new orthogonal 8-point approximate DCTs.
Third, we use a scaling method~\cite{Jridi2015} for proposing novel 16- and 32-point DCT approximations based on the optimal 8-point transforms, which are applicable to high-resolution image and video coding.
The best performing
transforms are assessed through
image, video, and hardware
experiments, and then compared with
state-of-the-art methods.
The results indicate their potential applicability in low-power or real-time video transmission scenarios~\cite{Harize2013,Ernawan2011,Potluri2014}.

The rest of this paper is organized as follows.
Section~\ref{sec:method} introduces our multiparametric
8-point DCT formulation as well as its underlying arithmetic complexity and fast algorithm.
Section~\ref{sec:Opt} presents a constrained  multicriteria optimization procedure for
acting over the search space associated with the proposed formalism and
introduces the optimal 8-point DCT-like transforms.
Section~\ref{sec:scaling} details
the adopted approach for transform scaling and presents
the resulting 16- and 32-point DCT approximations.
In Section~\ref{sec:experiments},
the optimal 8-point approximate  DCTs, and their scaled versions are submitted to image and video coding experiments in which
quality degradation is measured.
Section~\ref{sec:hardware} presents a hardware implementation of the optimal
8-point approximate DCTs.
A comprehensive comparison with competing methods is presented in
and in Sections~\ref{sec:Opt},
\ref{sec:scaling}, \ref{sec:experiments},
and \ref{sec:hardware}.
Section~\ref{sec:conclusions} concludes this work.

\section{Multiparametric approximate DCTs}
\label{sec:method}

Good approximations for the 8-point DCT matrix can be obtained
according to
a judicious
substitution
of the
64 matrix entries
by
values
in $\mathcal{C} = \{0,\pm \frac{1}{2},\pm 1, \pm2\}$
while
optimizing
a given
figure of merit.
Because the search space
increases quadratically,
approaching such optimization
problem
by exhaustive computational search
has proven to be
of limited practicality in contemporary computers~\cite{Potluri2013};
thus
motivating
alternative methods for obtaining approximate matrices.
A successful approach consists
of limiting the search space
according
to
constraints
from
a particular parametrization of
the DCT matrix
entries~\cite{Bouguezel2011,Tablada2015,Canterle2020,Coelho2018}.

In~\cite{Bouguezel2011},
a class of low-complexity transforms
was introduced
based
on a single free parameter.
A parametrization of the real entries of the Feig-Winograd fast algorithm~\cite{Feig1992} was proposed in~\cite{Tablada2015},
leading to a multiparametric class that encompasses several
DCT approximations.
At the same vein,
the Loeffler fast algorithm~\cite{Loeffler1989}
was given a parametrization
in~\cite{Coelho2018}
which furnished DCT approximations.
The methodology in~\cite{Canterle2020}
focused on deriving DCT approximations
based on
a unifying mathematical treatment
for
the suit of approximate transforms
introduced by Bouguezel, Ahmad, and Swamy~\cite{Bouguezel2011,Bouguezel2008,SaadBouguezel2009,
Bouguezel2013,Bouguezel2010}.

\subsection{Proposed Class of Low-complexity Transforms}\label{sec:PropClass}

We separate
transformation matrices
of the RDCT~\cite{Cintra2011} and the MRDCT~\cite{Bayer2012a}
and submit them
to a parametrization of its entries.
These approximations were selected because of
their relevance to the approximate DCT computation literature
in terms of combining
high coding performance
and
very low arithmetic complexity,
respectively.
Both are 8-point orthogonal transforms whose
associated low-complexity, integer transformation matrix entries are in the set $\{0, \pm 1\} \subset \mathcal{C}$, thus demanding
addition operations only.
Our multiparametric formulation accounts for element changes in MRDCT
matrix with respect to the RDCT.
The proposed multiparametric \emph{low-complexity} transformation matrix is given by
\begin{align}
\label{eq:parametric}
\mathbf{T(a)}=
\begin{bmatrix}
1 & 1 & 1 & 1 & 1 & 1 & 1 & 1\\
1 & a_1 & a_2 & 0 & 0 & -a_2 & -a_1 & -1\\
1 & 0 & 0 & -1 & -1 & 0 & 0 & 1\\
a_3 & 0 & -1 & -a_4 & a_4 & 1 & 0 & -a_3\\
1 & -1 & -1 & 1 & 1 & -1 & -1 & 1\\
a_5 & -1 & 0 & a_6 & -a_6 & 0 & 1 & -a_5\\
0 & -1 & 1 & 0 & 0 & 1 & -1 & 0\\
0 & -a_7 & a_8 & -1 & 1 & -a_8 & a_7 & 0
\end{bmatrix}
,
\end{align}
where
$\mathbf{a} = [a_1\enskip  a_2\enskip a_3\enskip a_4\enskip a_5\enskip a_6\enskip a_7\enskip a_8]^\top$ is the parameter vector.
In order to keep the complexity low and the search space
withing the available computation capabilities,
we focus on the case that the elements of $\mathbf{a}$
are in the set $\mathcal{C}= \{ 0, \pm\frac{1}{2}, \pm1 ,\pm2 \}$.
This set extends
the low-complexity sets
considered in~\cite{Cintra2011} and~\cite{Bayer2012a}
by the inclusion
of
$\{ \pm\frac{1}{2}, \pm2 \}$.
Thus,
the resulting parametrized matrices
might require bit-shifting operations.
The proposed multiparametric transform class in \eqref{eq:parametric} includes
$7^8 = 5,764,801$ different transformation matrices;
constituting
the search space
in which we aim at finding suitable transforms
for image and video coding.

\subsection{Fast Algorithm and Arithmetic Complexity} \label{sec:fastalg}

\begin{figure}
    \centering
    \includegraphics[width=.45\textwidth]{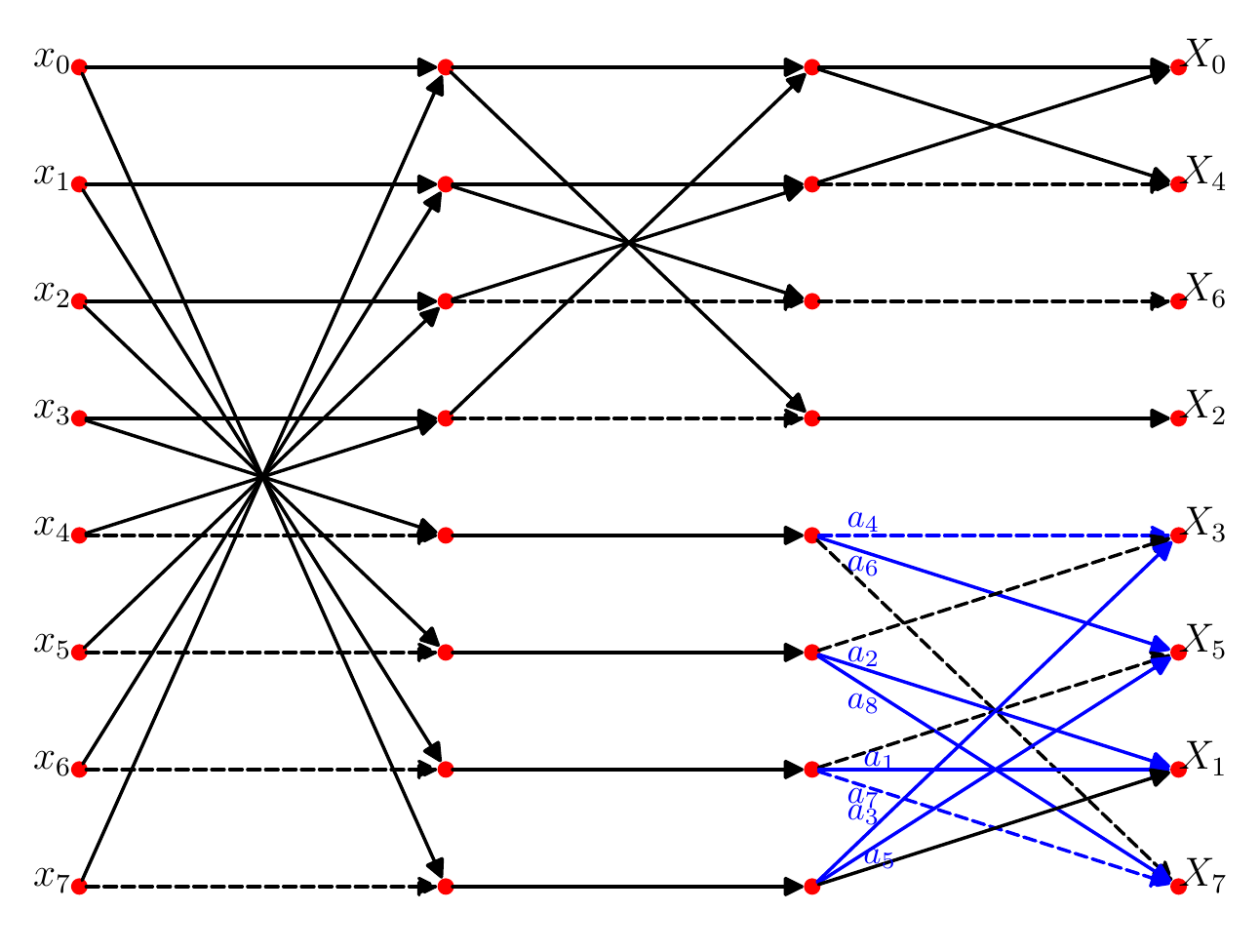}
    \caption{SFG for the proposed multiparametric transform. Solid and dashed arrows represent multiplication by 1 and -1, respectively. Blue arrows relate to the chosen parameter vector $\mathbf{a}$.}
    \label{fig:sfg}
\end{figure}

The arithmetic complexity provides a fair, unbiased assessment of the cost of applying a transform, and it is independent of the available technology~\cite{blahut2010fast,oppenheim1975digital}.
Directly multiplying the matrix
$\mathbf{T(a)}$
by a vector
is
as low as 24 additions
(for $\mathbf{a}$ is the null 8-point vector),
increasing according to the particular choices
of the parameter vector $\mathbf{a}$.
The computational cost of
the proposed multiparametric transform $\mathbf{T(a)}$
can be dramatically
reduced
by means of
the following
sparse factorization:
\begin{align}
\mathbf{T(a)} = \mathbf{P} \cdot \mathbf{K(a)} \cdot \mathbf{A_2} \cdot \mathbf{A_1},
\label{eq:fastalg}
\end{align}
where
\begin{align}
&\mathbf{A_1}=
\begin{bmatrix}
\mathbf{I}_4 & \bar{\mathbf{I}}_4 \\
\bar{\mathbf{I}}_4 & -\mathbf{I}_4
\end{bmatrix},
\quad
\mathbf{A_2}=
{\rm diag}\left(
\begin{bmatrix}
\mathbf{I}_2 & \bar{\mathbf{I}}_2 \\
\bar{\mathbf{I}}_2 & -\mathbf{I}_2
\end{bmatrix},
\mathbf{I}_4
\right),
\\
\footnotesize
&\mathbf{K(a)}=
{\rm diag}\left(
\begin{bmatrix}
1 & 1 \\
1 & -1
\end{bmatrix},-1,1,
\begin{bmatrix}
-a_4 & -1 & 0 & a_3 \\
a_6 & 0 & -1 & a_5 \\
0 & a_2 & a_1 & 1 \\
-1 & a_8 & -a_7 & 0
\end{bmatrix}
\right),
\end{align}
and
$\mathbf{P}$ is a permutation matrix, given by $(0)(1\,4\,3\,2\,6)(5)(7)$, in cyclic notation.
The matrices $\mathbf{I}_d$ and $\bar{\mathbf{I}}_d$ denote a identity and counter identity matrix of order $d$, respectively.
Matrices $\mathbf{A_1}$ and $\mathbf{A_2}$ require 8 and 4 additions, respectively.
Matrix $\mathbf{P}$ is a permutation matrix
with null arithmetic complexity.
The arithmetic cost of the matrix $\mathbf{K(a)}$
depends on the parameter vector $\mathbf{a}$
and
its
additive complexity ranges from
2 to 10~additions.
Fig.~\ref{fig:sfg} depicts the underlying signal flow graph~(SFG)
of the fast algorithm induced by~\eqref{eq:fastalg}.

The presented fast algorithm
computes $\mathbf{T(a)}$
at
additive and bit-shift complexities
given by
\begin{align}
\mathcal{A}(\mathbf{a}) = 22 - \sum^8_{i=1}\mathcal{I}_{\{ 0 \}}(a_i)
\;\text{and} \;
\mathcal{S}(\mathbf{a}) = \sum^8_{i=1}\mathcal{I}_{\left\lbrace  \frac{1}{2},2 \right\rbrace}(a_i),
\label{eq:arithmetic-complexity-shift}
\end{align}
where
$\mathcal{I}_{\Theta}(\theta)$
is the indicator function that returns $1$
if
$\theta \in \Theta$
and zero otherwise.

\section{Optimization on the Multiparametric Class}\label{sec:Opt}

To select which of the transforms are representative for the signal decorrelation problem, we set up a computational procedure that searches
for optimal transformations
according to preset objective metrics~\cite{Tablada2015,Canterle2020}.
Besides the orthogonality property, three
types of figures of merit are
considered
for rating a given approximate DCT:
(i) the arithmetic complexity for measuring its computational cost,
(ii)~proximity measures to the DCT in a euclidean distance sense,
and
(iii)~coding metrics for assessing its decorrelation capabilities.
By construction,
the proposed parametrization
has zero multiplicative cost.
Thus, for assessing the arithmetic complexity, we considered the total additive and bit-shift complexity given in
\eqref{eq:arithmetic-complexity-shift}.
For measuring the transform similarity to the DCT,
we adopted the
total error energy~\cite{Cintra2011}
and
the mean square error (MSE)~\cite{V.Britanak2006}.
The coding capabilities of a given approximate DCT are
often
measured
by the unified transform coding gain~\cite{Katto1991}
and
transform efficiency~\cite{V.Britanak2006} metrics.

\subsection{Orthogonality Constraint}\label{sec:OrtCons}

Approximate transforms are often desired to be orthogonal~\cite{V.Britanak2006}
because
orthogonality implies that both direct and inverse transforms
share the same computing structures~\cite{Ahmed1975}.
In fact, the inverse transform is obtained by transposition,
which
results in simpler
software and hardware
implementations~\cite{Jridi2015}.

In order to the matrix $\mathbf{T(a)}$
to posses an orthogonal vector basis,
the nondiagonal elements
of
$\mathbf{T(a)}\cdot\mathbf{T(a)}^\top$
must be null.
Thus, we have that
\begin{align}
    \mathbf{T(a)}\cdot\mathbf{T(a)}^\top=
    \begin{bmatrix}
8 & 0 & 0 & 0 & 0 & 0 & 0 & 0\\
0 & \tau_1 & 0 & \tau_2 & 0 & \tau_3 & 0 & \tau_4\\
0 & 0 & 4 & 0 & 0 & 0 & 0 & 0\\
0 & \tau_2 & 0 & \tau_5 & 0 & \tau_6 & 0 & \tau_7\\
0 & 0 & 0 & 0 & 8 & 0 & 0 & 0\\
0 & \tau_3 & 0 & \tau_6 & 0 & \tau_8 & 0 & \tau_9\\
0 & 0 & 0 & 0 & 0 & 0 & 4 & 0\\
0 & \tau_4 & 0 & \tau_7 & 0 & \tau_9 & 0 & \tau_{10}\\
    \end{bmatrix},
\end{align}
where
$\tau_1 = 2a_1^2+2a_2^2+2$,
$\tau_2 = -2a_2+2a_3$,
$\tau_3 = -2a_1+2a_5$,
$\tau_4 = -2a_1a_7+2a_2a_8$,
$\tau_5 = 2a_3^2+2a_4^2+2$,
$\tau_6 = 2a_3a_5-2a_4a_6$,
$\tau_7 = 2a_4-2a_8$,
$\tau_8 = 2a_5^2+2a_6^2+2$,
$\tau_9 = -2a_6+2a_7$,
 and
$\tau_{10} = 2a_7^2+2a_8^2+2$.
Thus,
the
orthogonality
conditions
are
\begin{gather}
\label{equations-condition-1}
\tau_2 = \tau_3 = \tau_4 = \tau_6 = \tau_7 = \tau_9 = 0
,
\\
\label{equations-condition-2}
\tau_1 \neq0, \tau_5 \neq0,
\tau_8 \neq0, \tau_{10} \neq0
.
\end{gather}
We obtain \emph{orthonormal} transforms $\mathbf{\tilde C(a)}$---referred
hereafter as DCT approximations---by means of
normalizing the energy of the basis vectors~\cite{Higham1986},
according to:
\begin{align}
\mathbf{\tilde C(a)} &= \mathbf{S(a)} \cdot \mathbf{T(a)},
\end{align}
where
$\mathbf{S(a)}$ is a diagonal matrix computed by
\begin{align}
\mathbf{S(a)} &=
\sqrt{\left(\mathbf{T(a)}\cdot\mathbf{T(a)}^\top\right)^{-1}},
\end{align}
and the square root operation is applied to each element of the argument matrix.
Relying on the orthogonality property, we have that
\begin{align}
 \mathbf{S(a)} &= \operatorname{diag}
 \left(
 \frac{1}{2\sqrt{2}},
 \frac{1}{\sqrt{\tau_1}},
 \frac{1}{2},
 \frac{1}{\sqrt{\tau_5}},
 \frac{1}{2\sqrt{2}},
 \frac{1}{\sqrt{\tau_8}},
 \frac{1}{2},
 \frac{1}{\sqrt{\tau_{10}}}
 \right).
\end{align}

If
\eqref{equations-condition-1}
and
\eqref{equations-condition-2}
are satisfied,
then
$
\tilde{\mathbf{C}}(\mathbf{a})
\cdot
\tilde{\mathbf{C}}(\mathbf{a})^\top
= \mathbf{I}_8
$.

\subsection{Arithmetic Complexity of the Approximation}\label{sec:ACA}

The arithmetic complexity associated to the approximate DCT $\mathbf{\tilde C(a)}$
can be regarded as identical
to the complexity of the
low-complexity matrix $\mathbf{T(a)}$,
because
the scaling factors of the diagonal matrix $\mathbf{S(a)}$
can be merged into the quantization step of
image and video enconders~\cite{Oliveira2018, Cintra2011, Bouguezel2010, SaadBouguezel2009, Potluri2013}.
Thus, the arithmetic complexity assessment presented
in
\eqref{eq:arithmetic-complexity-shift}
is applicable to both direct and inverse transforms,
because the same fast algorithm is applicable to both cases.

\subsection{Multicriteria Optimization}\label{sec:MultOpt}

By considering:
(i)~the search space implied by the proposed parametrization,
(ii)~the orthogonality condition,
and
(iii) the above discussed figures of merit,
we have
the following multicriteria minimization problem:
\begin{align}
\label{eq:optimization}
\begin{split}
\mathbf{a}_\text{opt}
=
\arg
\min_{\mathbf{a}}
\Big\{
\epsilon(\tilde{\mathbf{C}}(\mathbf{a})),
\operatorname{MSE}(\tilde{\mathbf{C}}(\mathbf{a})),
-C_g^*(\tilde{\mathbf{C}}(\mathbf{a})),
-\eta(\tilde{\mathbf{C}}(\mathbf{a})),
\mathcal{A}(\mathbf{a}),
\mathcal{S}(\mathbf{a})
 \Big\}
\end{split}
\end{align}
subject to
\begin{align}
\nonumber
\begin{cases}
a_i \in \mathcal{C}, \quad i = 1,2,\ldots,8%
\\
\tau_j = 0, \quad j = 2, 3, 4, 6, 7, 9%
\\
\tau_k \neq0, \quad k = 1, 5, 8, 10%
\end{cases},
\end{align}
where
$\epsilon(\cdot)$,
$\operatorname{MSE}(\cdot)$,
$C_g^*(\cdot)$,
and
$\eta(\cdot)$
compute the total error energy,
MSE,
unified transform coding gain,
and transform efficiency
of the argument transform.
Note that coding metrics are sought to be maximized, whereas proximity to the DCT (error) and complexity should be minimized.
The obtained optimal parameter vectors
$\mathbf{a}_\text{opt}$
readily
define
the optimal transforms
according to
$\tilde{\mathbf{C}}(\mathbf{a}_\text{opt})$.

\subsection{Optimal 8-point Transforms}
\label{sec:searching8pt}

The solution of
\eqref{eq:optimization} results on seven optimal 8-point transforms, which are listed in Table~\ref{T:E8}.
For simplicity, hereafter, we denote the optimal transforms as $\mathbf{\tilde{C}}_8^{(j)}$, where $j = 1,2,\ldots,7$.
Three of the optimal transformations
are state-of-the-art approximate DCTs
already archived in literature, namely:
the MRDCT~\cite{Bayer2012a},
the transform by Oliveira~\emph{et al.}~(OCBT)~\cite{Oliveira2013},
and the RDCT~\cite{Cintra2011}
($j = 1, 2, 6$,
respectively).
The transform OCBT can also be found in~\cite{Brahimi2020} and \cite{Canterle2020}.
However,
to the best of our knowledge,
the remaining four
transforms presented in Table~\ref{T:E8}
are novel contributions to the literature ($j = 3,4,5,7$).

\begin{table}
\caption{Optimal 8-point DCT approximations in the proposed class.} \label{T:E8}
\begin{center}
\begin{tabular}{ccccccc}
\hline
$j$ &	$\mathbf{a}_\text{opt}$ & Description \\
\hline
1 & [0	0	0	0	0	0	0	0]$^\top$ &	MRDCT~\cite{Bayer2012a}\\
2 & [1	0	0	0	1	0	0	0]$^\top$ &	OCBT~\cite{Oliveira2013} \\
3 & [1	0	0	1	1	0	0	1]$^\top$ &	New\\
4 & [1	0	0	0.5	1	0	0	0.5]$^\top$ &	New\\
5 & [1	1	1	-1	1	-1	-1	-1]$^\top$ &	New\\
6 & [1	1	1	1	1	1	1	1]$^\top$ &	RDCT~\cite{Cintra2011} \\
7 & [1	0.5	0.5	1	1	0.5	0.5	1]$^\top$ &	New\\
\hline
\end{tabular}
\end{center}
\end{table}

Table~\ref{T:Avaliacao8} presents
proximity and coding measurements relative to the DCT
along with
the arithmetic complexity associated to each of the obtained
optimal 8-point transforms.
Hereafter, in all the tables,
the best results are highlighted in boldface.
The approximation
$\mathbf{\tilde{C}}_{8}^{(1)}$ (MRDCT)
still holds the lowest additive cost
for an approximate DCT
requiring 14 additions only.
Also
the approximation
$\mathbf{\tilde{C}}_{8}^{(6)}$ (RDCT)
still maintain its status
as the transformation
with the the smallest total error energy measure.
On the other hand,
among the optimal transforms,
the proposed transform $\mathbf{\tilde{C}}_{8}^{(7)}$ possesses
the lowest MSE measurement,
a very low error energy, and attains the highest coding gain and efficiency.

\begin{table}[!h]
\caption{Measurements for the optimal 8-point approximate DCTs.}
\label{T:Avaliacao8}
\begin{center}
\begin{tabular}{ccccccc}
\hline
$j$ & $\epsilon(\cdot)$ &	MSE$(\cdot)$ &	$C_g^*(\cdot)$ &	$\eta(\cdot)$ & $\mathcal{A}(\cdot)$ & $\mathcal{S}(\cdot)$\\
\hline
1 & 8.6592 & 0.0594 & 7.3326 & 80.8969 & \textbf{14} & \textbf{0}\\
2 & 6.8543 & 0.0275 & 7.9118 & 85.6419 & 16 & \textbf{0}\\
3 & 5.0493 & 0.0246 & 7.9207 & 85.3793 & 18 & \textbf{0}\\
4 & 5.0184 & 0.0241 & 8.1102 & 86.8665 & 18 & 2\\
5 & 16.0260 & 0.0333 & 8.1571 & 88.1932 & 22 & \textbf{0}\\
6 & \textbf{1.7945} & 0.0098 & 8.1827 & 87.4297 & 22 & \textbf{0}\\
7 & 2.1443 & \textbf{0.0083} & \textbf{8.4261} & \textbf{89.1383} & 22 & 4\\
\hline
\end{tabular}
\end{center}
\end{table}

For comparison purposes,
we
compile
in Table~\ref{T:Comparacao8}
the performance measurements
of
representative competing transforms found in the literature:
the series of transforms introduced by Bouguezel, Ahmad and Swamy (BAS$_1$~\cite{Bouguezel2008},
BAS$_2$~\cite{SaadBouguezel2009},
BAS$_3$~\cite{Bouguezel2013},
BAS$_4$~\cite{Bouguezel2010}
and
BAS$_5$~\cite{Bouguezel2011});
the level 1 transform (LO)
by Lengwehasatit and Ortega~\cite{Lengwehasatit2004};
and
the approximations
reported
in~\cite{Tablada2015},~\cite{Oliveira2018}, and ~\cite{Canterle2020}
referred to as TBC, OCBSML, and CSBC,
respectively.
Such
selected
transforms
were chosen so that
they are not members
of the proposed class of approximations.
Hereafter, in the tables, the superscript
on CSBC and TBC
identifies the instance from the class of transforms reported in the original papers~\cite{Canterle2020,Tablada2015}; whereas the
one
next to BAS$_5$ indicates the quantity considered in the parametric formulation from~\cite{Bouguezel2011}.
The DCT and the integer DCT from HEVC~\cite{Meher2014} demand much more resources and are not used in the following comparisons.

\begin{table}
	\setlength{\tabcolsep}{3pt}
		\caption{Measurements for the competing 8-point approximate DCTs.} \label{T:Comparacao8}
		\begin{center}
			\begin{tabular}{lcccccc}
				\hline
				Transform & $\epsilon(\cdot)$ &	MSE$(\cdot)$ &	$C_g^*(\cdot)$ &	$\eta(\cdot)$ & $\mathcal{A}(\cdot)$ & $\mathcal{S}(\cdot)$\\
				\hline
                CSBC$^{(1)}$~\cite{Canterle2020} & 6.8543 & 0.0275 & 7.9118 & 85.6419 & \textbf{16} & \textbf{0} \\
                BAS$_5^{(0)}$~\cite{Bouguezel2011} & 26.8642 & 0.0710 & 7.9118 & 85.6419 & \textbf{16} & \textbf{0} \\
                BAS$_2$~\cite{SaadBouguezel2009} & 6.8543 & 0.0275 & 7.9126 & 85.3799 & 18 & \textbf{0} \\
                TBC$^{(6)}$~\cite{Tablada2015} & 8.6592 & 0.0588 & 7.3689 & 81.1788 & 18 & \textbf{0}\\
                BAS$_5^{(1)}$~\cite{Bouguezel2011} & 26.8642 & 0.0710 & 7.9126 & 85.3799 & 18 & \textbf{0} \\
                BAS$_1$~\cite{Bouguezel2008} & 5.9294 & 0.0238 & 8.1194 & 86.8626 & 18 & 2 \\
                BAS$_5^{(1/2)}$~\cite{Bouguezel2011} & 26.4018 & 0.0678 & 8.1194 & 86.8626 & 18 & 2\\
                CSBC$^{(9)}$~\cite{Canterle2020} & 4.1203 & 0.0214 & 8.1199 & 86.7297 & 20 & 3 \\
                TBC$^{(5)}$~\cite{Tablada2015} & 7.4138 & 0.0530 & 7.5753 & 83.0846 & 20 & 10 \\
                BAS$_3$~\cite{Bouguezel2013} & 35.0639 & 0.1023 & 7.9461 & 85.3138 & 24 & \textbf{0}\\
                LO~\cite{Lengwehasatit2004} & \textbf{0.8695} & 0.0061 & 8.3902 & 88.7023 & 24 & 2 \\
                BAS$_4$~\cite{Bouguezel2010} & 4.0935 & 0.0210 & 8.3251 & 88.2182 & 24 & 4 \\
                OCBSML~\cite{Oliveira2018} & 1.2194 & \textbf{0.0046} & \textbf{8.6337} & \textbf{90.4615} & 24 & 6 \\
				\hline
			\end{tabular}
		\end{center}
	\end{table}

To derive fair comparison, hereafter, we confront transforms possessing approximately the same arithmetic complexity.
The new approximate DCT $\mathbf{\tilde C}_8^{(3)}$ outperforms the other low-complexity transforms requiring 18 additions
in terms of error energy, MSE, and coding gain.
The proposed transform $\mathbf{\tilde C}_8^{(4)}$ achieves smaller total error energy and the higher efficiency if compared with other approaches that require exactly 18 addition and 2 bit-shifting operations.
The introduced transform $\mathbf{\tilde C}_8^{(5)}$ is more efficient than $\mathbf{\tilde C}_8^{(6)}$, but has smaller error energy, MSE and coding gain. No other listed transform demands exactly 22 additions.
The DCT approximation $\mathbf{\tilde C}_8^{(7)}$ has no direct competitor found.
Note, however, that $\mathbf{\tilde C}_8^{(7)}$ ranks in the third and fourth places in terms of MSE and total error energy, and attains the second place according to coding gain and efficiency measurements if compared with any other transform listed in Tables~\ref{T:Avaliacao8}~and~\ref{T:Comparacao8}.
Together with outstanding DCT-like transforms as $\mathbf{\tilde{C}}_{8}^{(6)}$ (RDCT)~\cite{Cintra2011}, LO~\cite{Lengwehasatit2004} and OCBSML~\cite{Oliveira2018}, the proposed transform $\mathbf{\tilde C}_8^{(7)}$ composes the new state-of-the-art on 8-point approximate DCTs.

\section{Scaling 8-point to 16- and 32-point Transforms} \label{sec:scaling}

Widely popular image and video coding standards---e.g. JPEG~\cite{Wallace1992}, H.262~\cite{Mitchell1996}, and H.264~\cite{Richardson2010}---employ
the 8-point DCT for decorrelation,
thus attracting the community efforts
to that specific length~\cite{Bayer2012a,Haweel2001,Bouguezel2011}.
More recently,
the
HEVC standard
proposed to use
DCT-like transforms of different lengths: $4$, $8$, $16$, and $32$~\cite{Ohm2012}.
Such design choice enhances the compression of high-resolution video and improves
the coding efficiency mainly at low bit-rates~\cite{Wien2014}.
Generally, small-sized transforms cope with textured regions, whereas the large-sized ones act on smoother video content~\cite{MahsaT.Pourazard2012}.

Therefore,
there is a demand for  low-cost DCT-like transforms,
since the computational complexity of the exact DCT grows non-linearly~\cite{Jridi2015}.
Nevertheless,
there are only a few works focusing on
natively approximating the 16- or 32-point
DCT~\cite{DaSilveira2016a,DaSilveira2017a,Bayer2012}.
A practical approach
for scaling up 8-point transforms to larger transforms
of size 16 and 32
consists
of applying the method proposed by
Jridi, Alfalou, and Meher (JAM)~\cite{Jridi2015}.
In a nutshell, the JAM method takes two instances of a low-complexity multiplierless transform of length $N$ to compose another transform of length $2N$.
The total arithmetic complexity of the $2N$-point transform is kept low.
It requires twice the number of bit-shifts and twice plus $2N$ additions demanded by the original $N$-point transform.
Originally,
the RDCT ($\mathbf{\tilde C}_8^{(6)}$) was used in the scaling process introduced in~\cite{Jridi2015}.
For brevity, we refer the reader to the original paper for more details about the scalable JAM method~\cite{Jridi2015}.

\subsection{16-point Scaled Transforms}

The proposed optimal 8-point transforms
were submitted to the JAM method
in order to generate scaled transforms of length 16.
Table~\ref{T:Avaliacao16} summarizes the quality and complexity measurements for these DCT approximations.
To the best of our knowledge, five from the seven derived 16-point
transforms are novel contributions to the literature.
The 16-point transforms $\mathbf{\tilde C}_{16}^{(6)}$ and $\mathbf{\tilde C}_{16}^{(2)}$ were introduced in (JAM)~\cite{Jridi2015} and (CSBC$^{(1)}$)~\cite{Canterle2020}, respectively.
The results
from
Table~\ref{T:Avaliacao8}
and
Table~\ref{T:Avaliacao16}
shows
that
the JAM scaling
could
roughly
transfer the performance
from
the 8-point
to
the 16-point approximations.

\begin{table}[!h]
\caption{Measurements for the scaled 16-point approximate DCTs.} \label{T:Avaliacao16}
\begin{center}
\begin{tabular}{ccccccc}
\hline
$j$ & $\epsilon(\cdot)$ &	MSE$(\cdot)$ &	$C_g^*(\cdot)$ &	$\eta(\cdot)$ & $\mathcal{A}(\cdot)$ & $\mathcal{S}(\cdot)$\\
\hline
1 & 29.7486 & 0.0935 & 7.5816 & 66.0681 & \textbf{44} & \textbf{0}\\
2 & 25.1300 & 0.0674 & 8.1577 & 70.9808 & 48 & \textbf{0}\\
3 & 21.5172 & 0.0646 & 8.1664 & 70.5897 & 52 & \textbf{0}\\
4 & 21.6809 & 0.0644 & 8.3560 & 72.1975 & 52 & 4\\
5 & 41.1430 & 0.0707 & 8.4036 & 73.8217 & 60 & \textbf{0}\\
6 & \textbf{14.7402} &  \textbf{0.0506} & 8.4285 & 72.2296 & 60 & \textbf{0}\\
7 & 15.8124 & 0.0507 & \textbf{8.6711} & \textbf{75.8460} & 60 & 8\\
\hline
\end{tabular}
\end{center}
\end{table}

Table~\ref{T:Comparacao16}
lists
representative
16-point DCT approximations:
the
BAS$_{3}$ and BAS$_{4}$ transforms,
the approximations in~\cite{DaSilveira2016a} (SOBCM)
and~\cite{DaSilveira2017a} (SBCKMK);
and
the approximation in~\cite{Bayer2012} (BCEM).
We
also included
JAM-scaled versions of the
8-point approximations CSBC~\cite{Canterle2020} and OCBSML~\cite{Oliveira2018}.

\begin{table}[!h]
	\setlength{\tabcolsep}{2.4pt}
		\caption{
		Measurements for competing 16-point approximate DCTs.
		} \label{T:Comparacao16}
		\begin{center}
			\begin{tabular}{lcccccc}
				\hline
				Transform & $\epsilon(\cdot)$ &	MSE$(\cdot)$ &	$C_g^*(\cdot)$ &	$\eta(\cdot)$ & $\mathcal{A}(\cdot)$ & $\mathcal{S}(\cdot)$\\
				\hline
				SOBCM~\cite{DaSilveira2016a} & 40.9996 & 0.0947 & 7.8573 & 67.6078 & \textbf{44} & \textbf{0} \\
				CSBC$^{(4)}$~\cite{Canterle2020} & 20.8777 & 0.0648 & 8.1587 & 71.4837 & 52 & 2 \\
				CSBC$^{(5)}$~\cite{Canterle2020} & 23.0211 & 0.0641 & 8.3653 & 71.8269 & 52 & 4\\
				CSBC$^{(9)}$~\cite{Canterle2020}& 18.7688 & 0.0615 & 8.3663 & 72.3414 & 56 & 6 \\
				CSBC$^{(10)}$~\cite{Canterle2020} & 19.6427 & 0.0621 & 8.3659 & 72.1040 & 56 & 6 \\
				SBCKMK~\cite{DaSilveira2017a} & 30.3230 & 0.0639 & 8.2950 & 70.8315 & 60 & \textbf{0} \\
				CSBC$^{(13)}$~\cite{Canterle2020} & 18.5159 & 0.0599 & 8.3647 & 72.6288 & 60 & 4 \\
				BAS$_{3}$~\cite{Bouguezel2013} & 97.8678 & 0.4520 & 8.1941 & 70.6465 & 64 & \textbf{0} \\
				BAS$_{4}$~\cite{Bouguezel2010} & 16.4071 & 0.0564 & 8.5208 & 73.6345 & 64 & 8 \\
				OCBSML~\cite{Oliveira2018} & 13.7035 & 0.0474 & \textbf{8.8787} & \textbf{76.8108} & 64 & 12 \\
				BCEM~\cite{Bayer2012} & \textbf{8.0806} & \textbf{0.0465} & 7.8401 & 65.2789 & 72 & \textbf{0} \\
				\hline
			\end{tabular}
		\end{center}
	\end{table}

The proposed approximation
$\mathbf{\tilde C}_{16}^{(1)}$ outperforms SOBCM in proximity metrics, requiring 44 additions only.
Transform $\mathbf{\tilde C}_{16}^{(3)}$ has no direct competitor with the same arithmetic cost but performs close to CSBC$^{(4)}$
and CSBC$^{(5)}$, which require two and four extra bit-shifting operations, respectively.
The DCT approximation $\mathbf{\tilde C}_{16}^{(4)}$ attains smaller error energy and higher efficiency when compared with CSBC$^{(5)}$.
The proposed approximate DCT $\mathbf{\tilde C}_{16}^{(5)}$ outperforms SBCKMK in coding metrics, and achieves better transform efficiency than $\mathbf{\tilde C}_{16}^{(6)}$.
The transform $\mathbf{\tilde C}_{16}^{(7)}$ ranks on the fourth place in terms of proximity to the DCT, and is the second best-performing in coding metrics among the transforms in Tables~\ref{T:Avaliacao16} and \ref{T:Comparacao16}.

\subsection{32-point Scaled Transforms}

By invoking the JAM method twice,
we can scale 8-point transforms and obtain
32-point DCT approximations.
The obtained error and coding measurements as well as the arithmetic complexity of the obtained
32-point transforms are presented in Table~\ref{T:Avaliacao32}.
We found $\mathbf{\tilde C}_{32}^{(1)}$, $\mathbf{\tilde C}_{32}^{(3)}$, $\mathbf{\tilde C}_{32}^{(4)}$, $\mathbf{\tilde C}_{32}^{(5)}$, and $\mathbf{\tilde C}_{32}^{(7)}$ as contributions to the literature.
The scaled transforms $\mathbf{\tilde C}_{32}^{(2)}$ and $\mathbf{\tilde C}_{32}^{(6)}$ coincide with 32-point versions of CSBC$^{(1)}$ and JAM proposal, respectively.
Peering approaches and their respective quality and cost measurements are listed in Table~\ref{T:Comparacao32}.
Excepting for SOBCM, SBCKMK, and BCEM which are 16-point transforms only, the other competing approaches are the 32-point versions of those in Table~\ref{T:Comparacao16}. Namely, we consider the 32-point BAS$_{3}$, BAS$_{4}$, CSBC and OCBSML transforms.

\begin{table}[!h]
\caption{Measurements for the scaled 32-point approximate DCTs.} \label{T:Avaliacao32}
\begin{center}
\begin{tabular}{ccccccc}
\hline
$j$ & $\epsilon(\cdot)$ &	MSE$(\cdot)$ &	$C_g^*(\cdot)$ &	$\eta(\cdot)$ & $\mathcal{A}(\cdot)$ & $\mathcal{S}(\cdot)$\\
\hline
1 & 77.7215 & 0.1497 & 7.6584 & 52.2784 & \textbf{120} & \textbf{0}\\
2 & 68.1287&  0.1278 & 8.2306 & 56.1785 & 128 & \textbf{0}\\
3 & 61.2029 & 0.1251 & 8.2393 & 55.8320 & 136 & \textbf{0}\\
4 & 61.7212 & 0.1252 & 8.4287 & 57.1200 & 136 & 8\\
5 & 96.7291 & 0.1302 & 8.4771 & 58.4748 & 152 & \textbf{0}\\
6 & \textbf{48.0956} & \textbf{0.1124} & 8.5010 & 56.9700 & 152 & \textbf{0}\\
7 & 50.4638 & 0.1133 & \textbf{8.7429} & \textbf{60.4018} & 152 & 16\\
\hline
\end{tabular}
\end{center}
\end{table}

\begin{table}[!h]
		\caption{
		Measurements for competing 32-point approximate DCTs.\label{T:Comparacao32}
		}
		\begin{center}
			\begin{tabular}{l@{\enskip \,\,}c@{\enskip \,\,}c@{\enskip \,\,}c@{\enskip \,\,}c@{\enskip \,\,}c@{\enskip \,\,}c}
				\hline
				Transform & $\epsilon(\cdot)$ &	MSE$(\cdot)$ &	$C_g^*(\cdot)$ &	$\eta(\cdot)$ & $\mathcal{A}(\cdot)$ & $\mathcal{S}(\cdot)$\\
				\hline
				CSBC$^{(4)}$~\cite{Canterle2020} & 59.4743 & 0.1253 & 8.2320 & 56.7808 & \textbf{136} & 4 \\
				CSBC$^{(5)}$~\cite{Canterle2020} & 63.9307 & 0.1249 & 8.4382 & 56.7210 & \textbf{136} & 8\\
				CSBC$^{(6)}$~\cite{Canterle2020} & 60.6931 & 0.1218 & 8.2516 & 56.4665 & 144 & \textbf{0}\\
				CSBC$^{(9)}$~\cite{Canterle2020} & 55.2764 & 0.1224 & 8.4396 & 57.3346 & 144 & 12 \\
				CSBC$^{(10)}$~\cite{Canterle2020} & 56.3736 & 0.1227 & 8.4390 & 56.9787 & 144 & 12 \\
				CSBC$^{(13)}$~\cite{Canterle2020} & 52.9321 & 0.1186 & 8.4389 & 57.5669 & 152 & 8 \\
				BAS$_{3}$~\cite{Bouguezel2013} & 192.1804 & 0.7609 & 8.2693 & 55.9114 & 160 & \textbf{0} \\
				BAS$_{4}$~\cite{Bouguezel2010} & 117.0653 & 0.2411 & 8.4998 & 58.4956 & 160 & 16 \\
				OCBSML~\cite{Oliveira2018} &  \textbf{46.2658} & \textbf{0.1104} & \textbf{8.9505} & \textbf{61.0272} & 160 & 24 \\
                \hline
			\end{tabular}
		\end{center}
	\end{table}

To the best of our knowledge, the novel transform $\mathbf{\tilde C}_{32}^{(1)}$ possesses the smaller arithmetic complexity archived in the literature.
Transform $\mathbf{\tilde C}_{32}^{(3)}$ has smaller error energy than CSBC$^{(5)}$, smaller MSE
than CSBC$^{(4)}$, and higher coding gain than CSBC$^{(4)}$,
while
requiring fewer arithmetic operations.
The transform
$\mathbf{\tilde C}_{32}^{(4)}$ outperforms CSBC$^{(5)}$ in terms of error energy and efficiency, both requiring 136 additions and 8 bit-shifts.
$\mathbf{\tilde C}_{32}^{(5)}$ has no direct competitors, but presents competitive coding gain and transform efficiency measurements.
Once again, $\mathbf{\tilde C}_{32}^{(5)}$ stands out, by ranking on third and second places for proximity to the exact DCT and coding capabilities, respectively, among all considered 32-point transforms.

\section{Application to Image and Video Coding} \label{sec:experiments}

In this section,
we describe
computational experiments
on both still-image and video compression
aiming at assessing
the
behavior of the selected transforms on such applications.

\subsection{Still-Image Compression} \label{subsec:image}

We performed a JPEG-like experiment to
assess
the optimal 8-point DCT-like transforms and their scaled 16- and 32-point counterparts
in the context of still-image compression.
The experiment consisted of subdividing the input image into disjoint blocks $\mathbf{A}$ of size $N\times N$.
Each block was individually processed as follows.
The direct transformation was applied to $\mathbf{A}$ according to $\mathbf{B} = \mathbf{T}_N \cdot \mathbf{A} \cdot \mathbf{T}_N^{\top}$, where $\mathbf{T}_N$ is a $N$-point orthogonal transform and $\mathbf{B}$ is the resulting transformed block.
Using the zig-zag scan sequence~\cite{Wallace1992}, we kept the first $r$ coefficients while setting the remaining coefficients to zero.
The truncated
block is represented by $\mathbf{\tilde B}$.
Then,
we applied the inverse transformation to each block $\mathbf{\tilde B}$ through $\mathbf{\tilde A} = \mathbf{T}^{\top}_N \cdot \mathbf{\tilde B} \cdot \mathbf{T}_N$.
The correct rearrangement of all the blocks $\mathbf{\tilde A}$ resulted on a $\mathbf{T}_N$-compressed version of the input image at compression rate $r/N^2$.

In this experiment, we
evaluated the performance of a given transform $\mathbf{T}_N$ by objectively assessing the quality of the compressed image.
For that end, we employed the peak signal-to-noise-ratio (PSNR)~\cite{Huynh-Thu2008} and the  structural similarity index (SSIM)~\cite{Wang2004}
following the procedure described
in~\cite{Brahimi2020, Jridi2015, Haweel2001, Bouguezel2011, Bouguezel2008, SaadBouguezel2009}.
We also include results for the learned perceptual image patch similarity (LPIPS)~\cite{zhang2018unreasonable}, a
perceptual-based metric, which correlates to the mean opinion score~\cite{Khrulkov_2021_CVPR}.
We report the averaged results for  45 $512 \times 512$ grayscale images from a public database~\cite{uscsipi}, with $r$ varying from $1$ to approximately $0.75N^2$, with $N = 8, 16, 32$.
The chosen quantities of retained coefficients $r$ roughly correspond to compression rates ranging from $25\%$ to $99\%$.

\subsubsection{Results for the 8-point Approximations}

\begin{figure*}[!h]
\begin{center}
\subfigure[Average SSIM]{
{\includegraphics[height=0.30\textwidth]{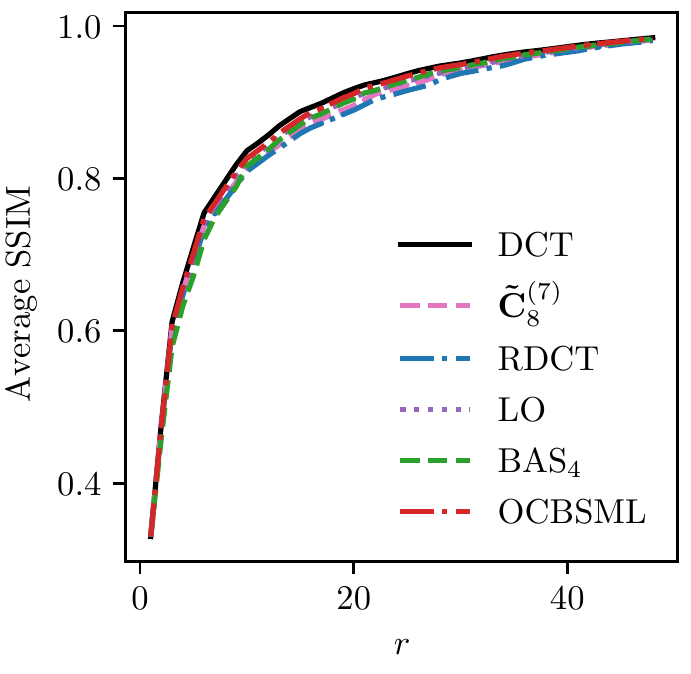}}\label{F:MSSIM}}
\,
\subfigure[Average PSNR]{
{\includegraphics[height=0.30\textwidth]{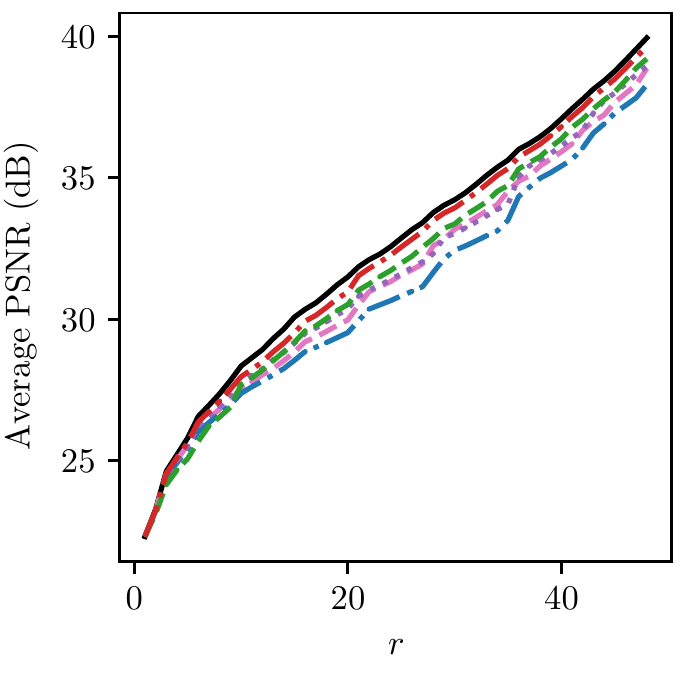}}\label{F:PSNR}}
\,
\subfigure[Average LPIPS]{
{\includegraphics[height=0.30\textwidth]{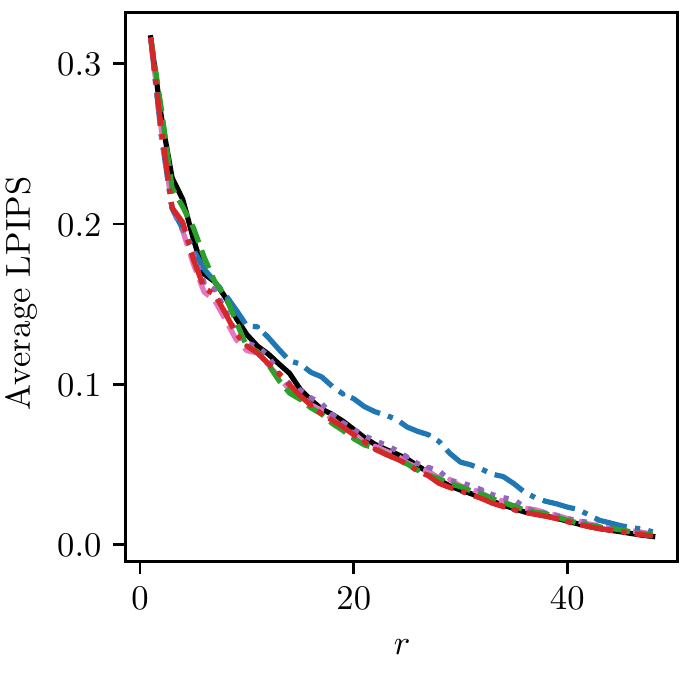}}\label{F:LPIPS}}
\,
\subfigure[Average SSIM / $\mathcal{A}(\cdot)$]{
{\includegraphics[height=0.3\textwidth]{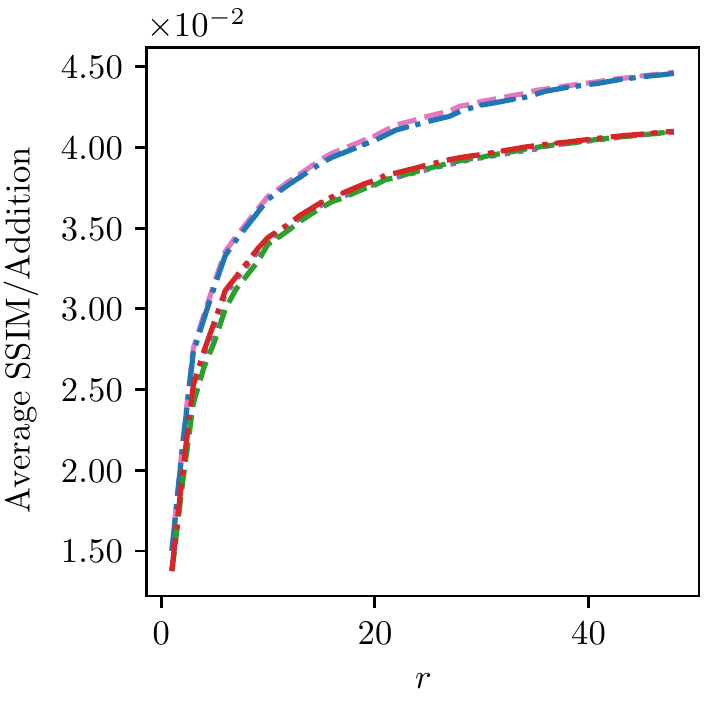}}\label{F:MSSIMadd}}
\,
\subfigure[Average PSNR / $\mathcal{A}(\cdot)$]{
{\includegraphics[height=0.3\textwidth]{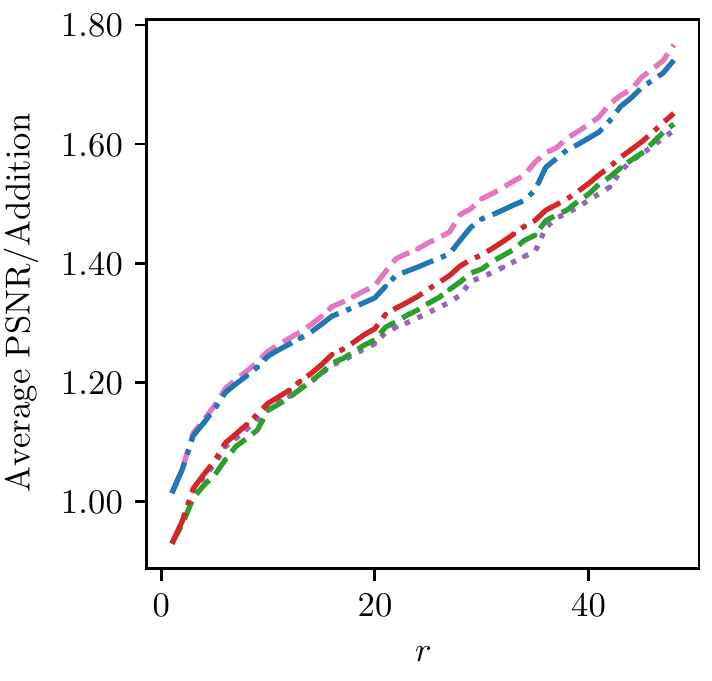}}\label{F:PSNRadd}}
\,
\subfigure[Average LPIPS / $\mathcal{A}(\cdot)$]{
{\includegraphics[height=0.3\textwidth]{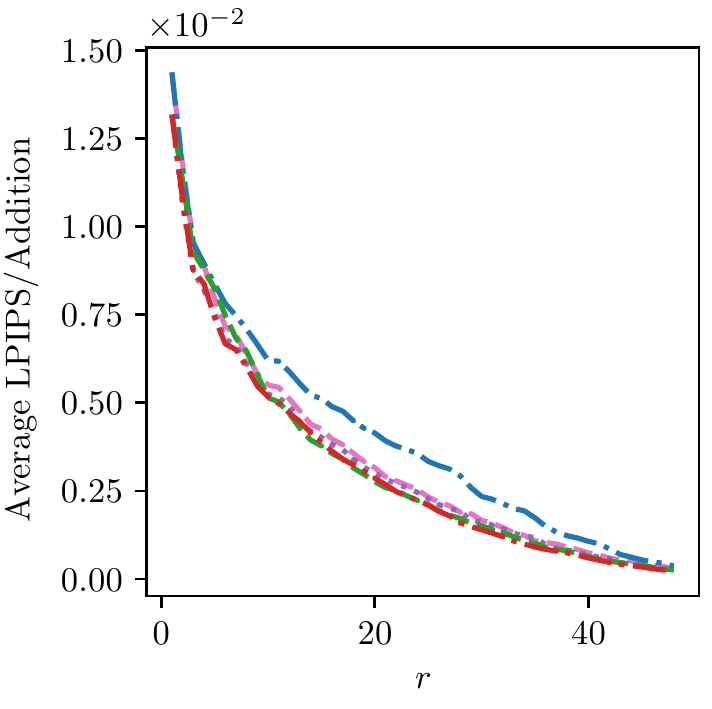}}\label{F:LPIPSadd}}
\caption{Image compression results for novel and competing 8-point DCT approximations. The legends are shared across the multiple graphics.}
\label{F:compcompre}
\end{center}
\end{figure*}

Fig.~\ref{F:MSSIM} and Fig.~\ref{F:PSNR} depict the average SSIM and PSNR measurements, respectively, for selected 8-point transforms.
Namely, we separated
the best-performing novel transform $\mathbf{\tilde C}^{(7)}_8$, and
the followings competing methods:
LO, OCBSML, RDCT, BAS$_4$, and the exact DCT.
Note that the transform $\mathbf{\tilde C}^{(7)}_8$ achieves the second-best PSNR gains for high compression rates ($r < 5$).
It also performs comparably to the LO approximation
for intermediary compression rates ($20 < r < 35$), and outperforms the RDCT in all the cases.
Similar results are obtained
for the SSIM measurements.
Fig.~\ref{F:LPIPS} depicts the LPIPS measurements, suggesting that approximate transforms can outperform the DCT for some $r$ values.

Complementary, we provide the SSIM, PSNR and LPIPS gains per addition operation---which
can be useful
for comparing transforms of different complexities.
Bit-shifting operations are often regarded as virtually costless in hardware implementations~\cite{Oliveira2017, Oliveira2018},
 and are thus suppressed in our analysis.
Fig.~\ref{F:MSSIMadd} and Fig.~\ref{F:PSNRadd} show that the proposed DCT-like transform $\mathbf{\tilde C}^{(7)}_8$ has consistently higher SSIM gains per addition unit, also reflected in terms of PSNR measurements.
The RDCT, well-known for its outstanding coding capabilities, behaves similarly to $\mathbf{\tilde C}^{(7)}_8$, but presenting slightly worse results.
Fig.~\ref{F:LPIPSadd} shows the LPIPS values per addition, where $\mathbf{\tilde C}^{(7)}_8$ roughly compares to LO, BAS$_4$, and OCBSML.

\subsubsection{Results for the 16-point Approximations}

\begin{figure*}[!h]
\begin{center}
\subfigure[Average SSIM]{
{\includegraphics[height=0.3\textwidth]{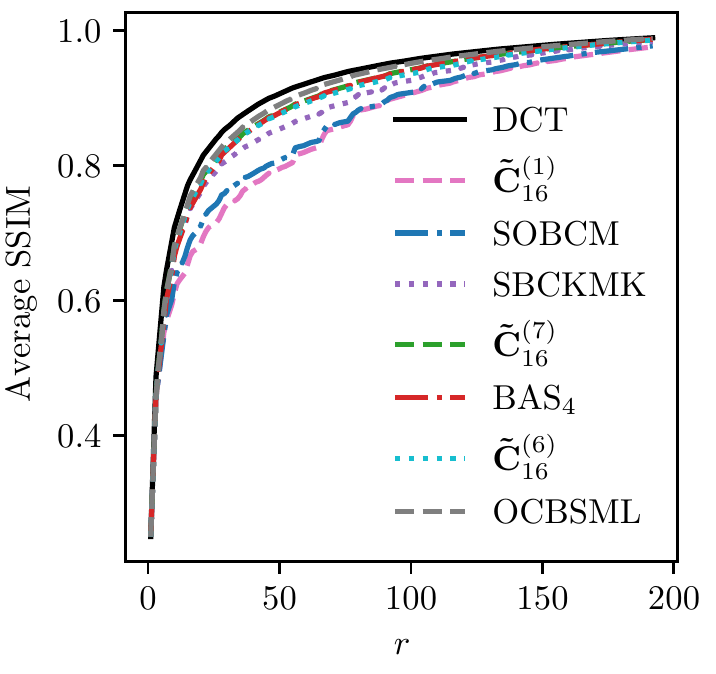}}\label{F:MSSIM16}}
\,
\subfigure[Average PSNR]{

{\includegraphics[height=0.3\textwidth]{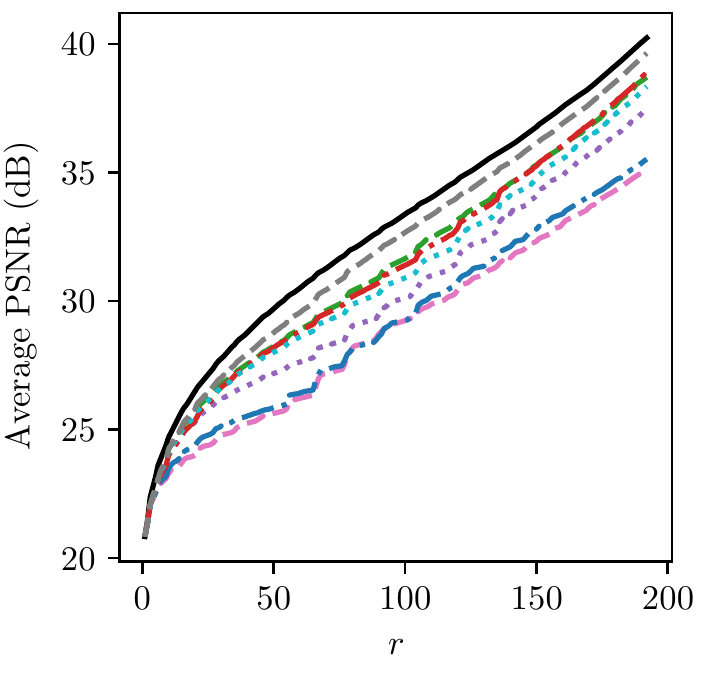}}\label{F:PSNR16}}
\,
\subfigure[Average LPIPS]{
{\includegraphics[height=0.3\textwidth]{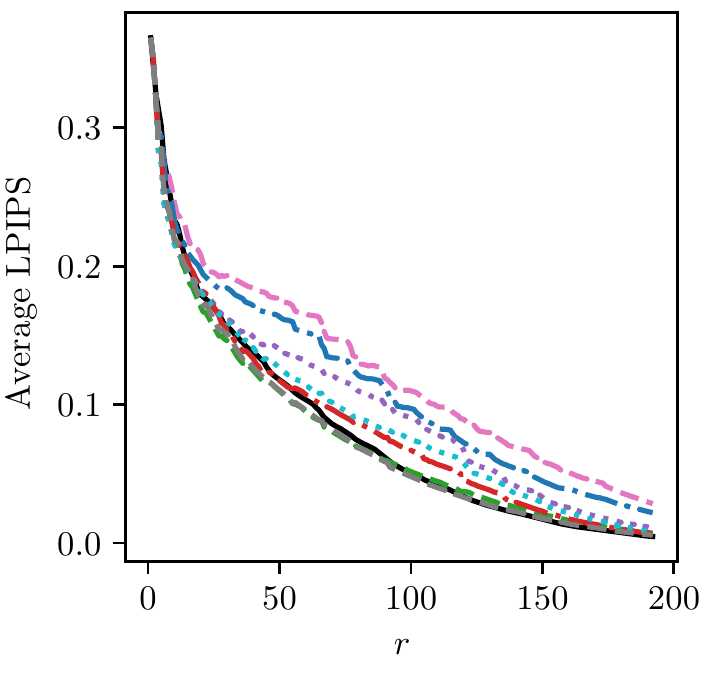}}\label{F:LPIPS16}}
\,
\subfigure[Average SSIM / $\mathcal{A}(\cdot)$]{
{\includegraphics[height=0.3\textwidth]{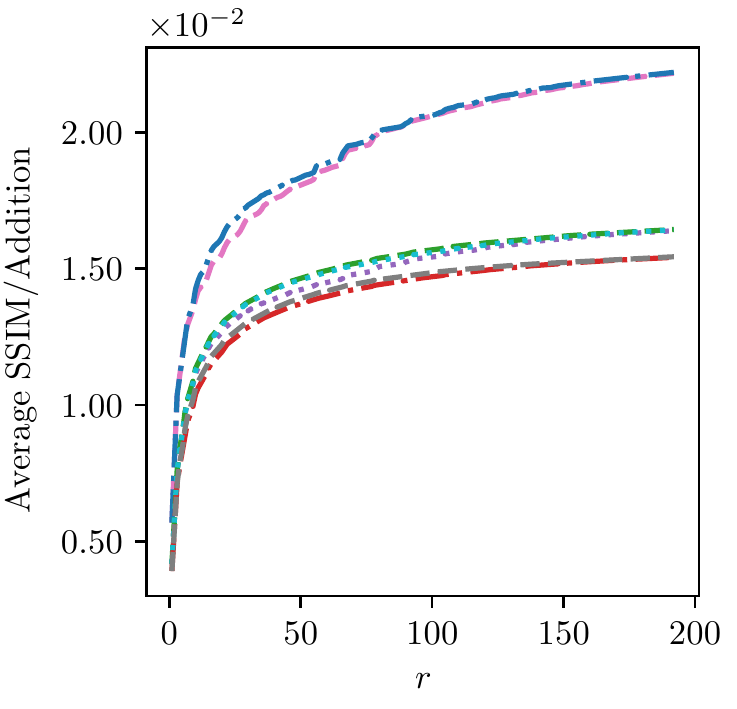}}
\label{F:MSSIMadd16}}
\,
\subfigure[Average PSNR / $\mathcal{A}(\cdot)$]{
{\includegraphics[height=0.3\textwidth]{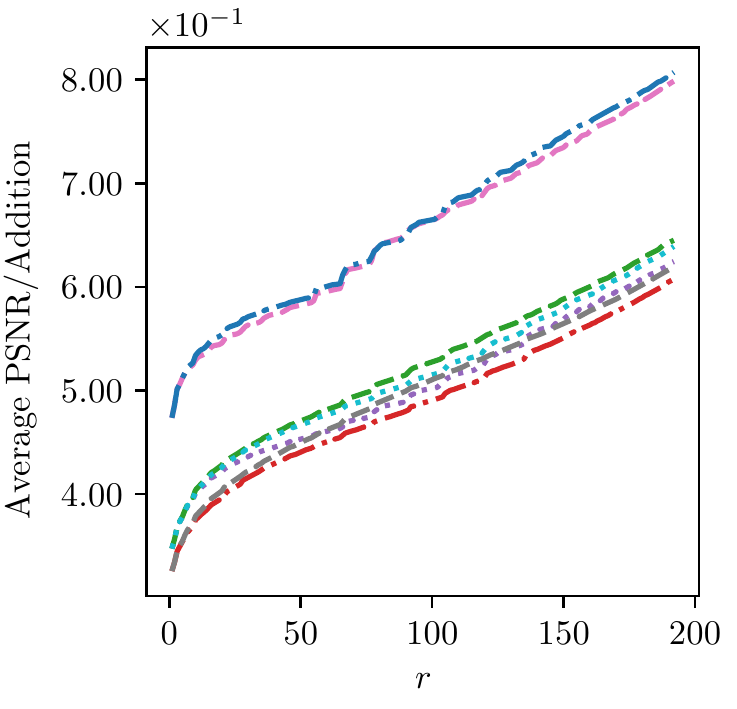}}
\label{F:PSNRadd16}}
\,
\subfigure[Average LPIPS / $\mathcal{A}(\cdot)$]{
{\includegraphics[height=0.3\textwidth]{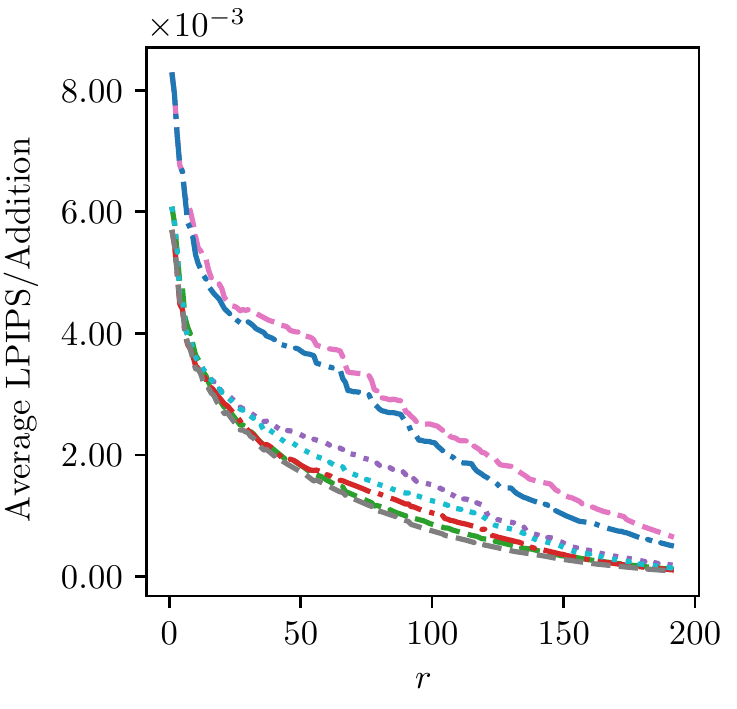}}\label{F:LPIPSadd16}}
\caption{Image compression results for proposed and competing 16-point DCT approximations. The legends are shared across the multiple graphics.}
\label{F:compcompre16}
\end{center}
\end{figure*}

Fig.~\ref{F:MSSIM16} and Fig.~\ref{F:PSNR16} present the average SSIM and PSNR measurements for the experiment involving 16-point transforms.
We separated the scaled transforms $\mathbf{\tilde C}^{(1)}_{16}$, $\mathbf{\tilde C}^{(7)}_{16}$, and $\mathbf{\tilde C}^{(6)}_{16}$, besides peering approaches like BAS$_4$, SBCKMK, SOBCM, OCBSML, and the exact DCT.
The proposed transform $\mathbf{\tilde C}^{(7)}_{16}$ outperforms all other approximate DCTs, excepting for OCBSML, for $r < 140$.
The transform $\mathbf{\tilde C}^{(1)}_{16}$ achieves comparable PSNR and SSIM measurements roughly for $50 < r < 100$ if compared with SOBCM, both requiring 44 additions only.
Fig.~\ref{F:LPIPS16} show the LPIPS scores. The results indicate $\mathbf{\tilde C}^{(7)}_{16}$ and OCBSML as best-performing together with the exact DCT.

The SSIM, PSNR, and LPIPS curves normalized by the number of additions required by the selected low-complexity 16-point transforms are shown in Fig.~\ref{F:MSSIMadd16},  Fig.~\ref{F:PSNRadd16},  and Fig.~\ref{F:LPIPSadd16}, respectively.
The curves
indicate a favorable performance
for $\mathbf{\tilde C}^{(1)}_{16}$
in terms of SSIM/PSNR gain per addition unit.
The approximate DCT $\mathbf{\tilde C}^{(7)}_{16}$ has comparable SSIM gain per addition regarding $\mathbf{\tilde C}^{(6)}_{16}$, and consistently outperforms SBCKMK, BAS$_4$ and OCBSML both in normalized PSNR and SSIM gains per addition operation.

\subsubsection{Results for the 32-point Approximations}

\begin{figure*}[!h]
\begin{center}
\subfigure[Average SSIM]{
{\includegraphics[height=0.3\textwidth]{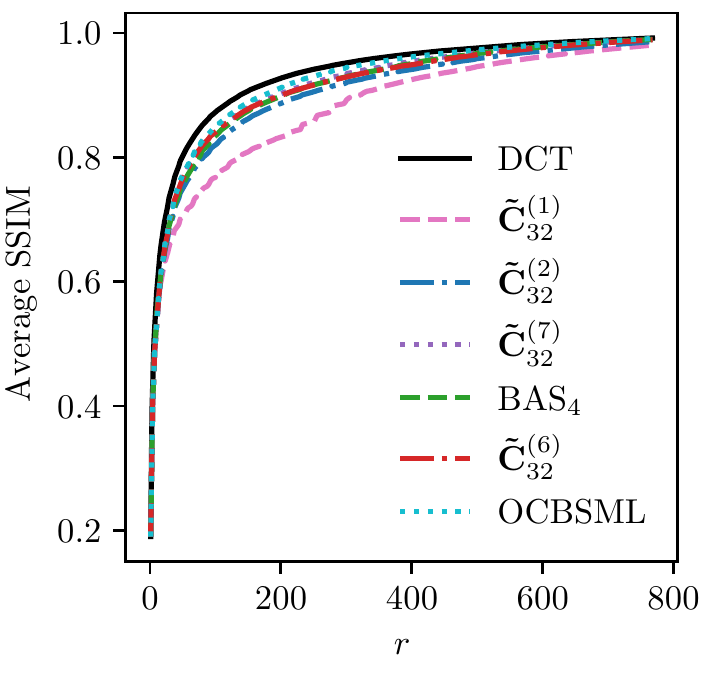}}\label{F:MSSIM32}}
\,
\subfigure[Average PSNR]{
{\includegraphics[height=0.3\textwidth]{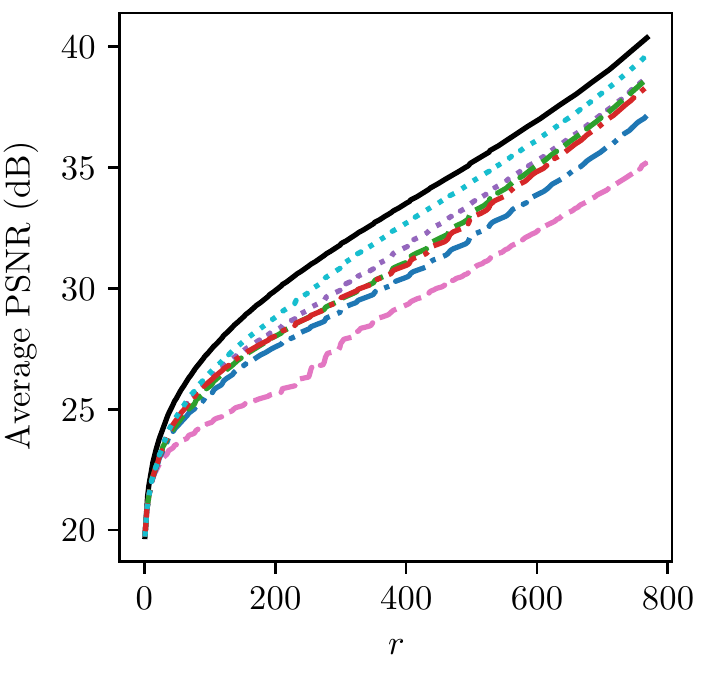}}\label{F:PSNR32}}
\,
\subfigure[Average LPIPS]{
{\includegraphics[height=0.3\textwidth]{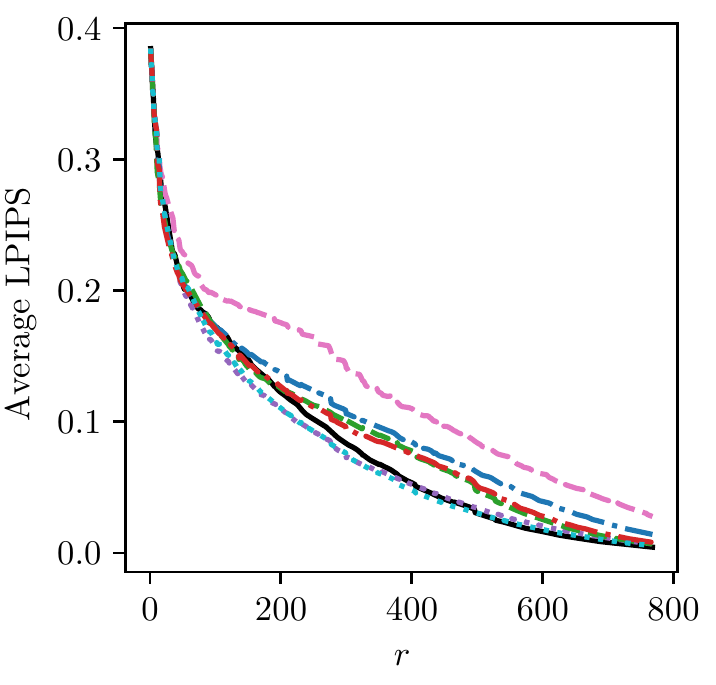}}\label{F:LPIPS32}}
\,
\subfigure[Average SSIM / $\mathcal{A}(\cdot)$]{
{\includegraphics[height=0.3\textwidth]{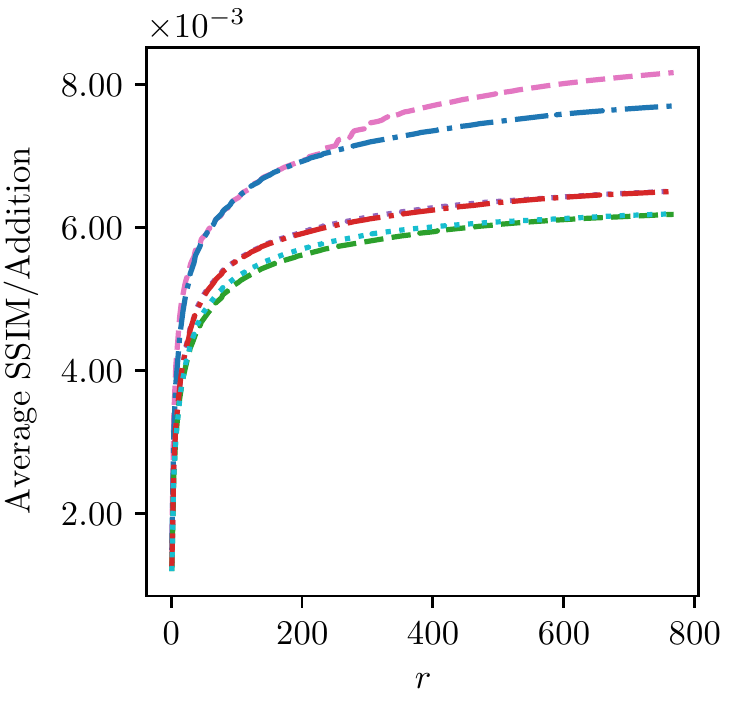}}\label{F:MSSIMadd32}}
\,
\subfigure[Average PSNR / $\mathcal{A}(\cdot)$]{
{\includegraphics[height=0.3\textwidth]{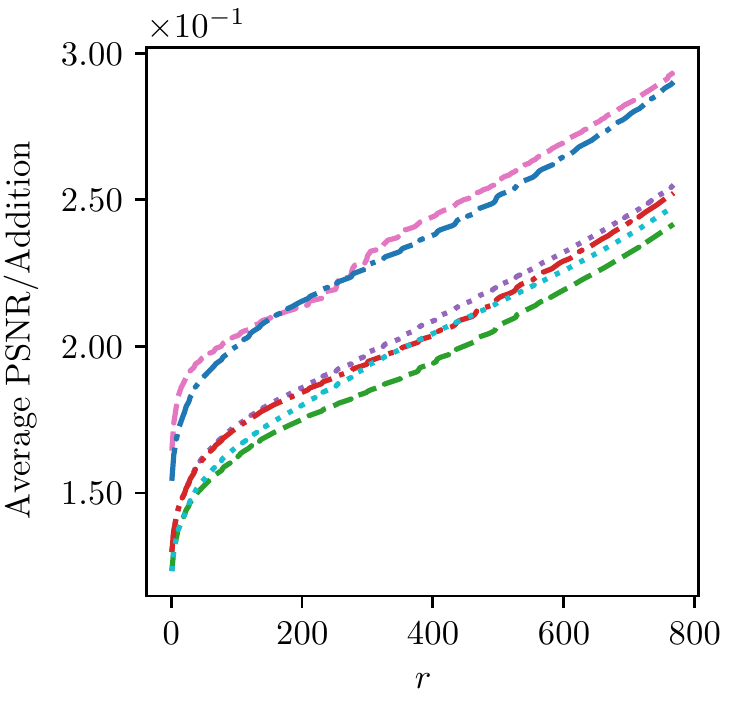}}
\label{F:PSNRadd32}}
\,
\subfigure[Average LPIPS / $\mathcal{A}(\cdot)$]{
{\includegraphics[height=0.3\textwidth]{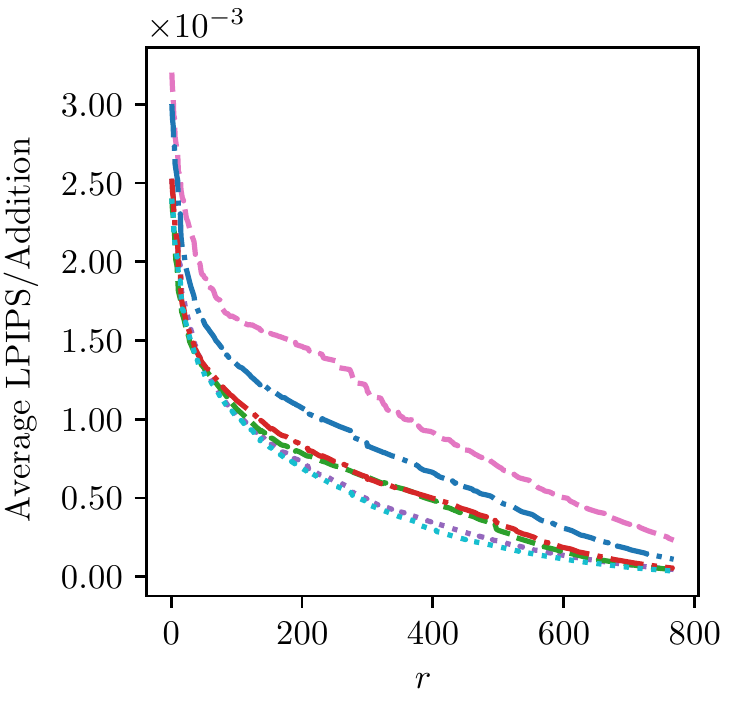}}\label{F:LPIPSadd32}}
\caption{Image compression results for the proposed and competing 32-point DCT approximations. The legends are shared across the multiple graphics.}
\label{F:compcompre32}
\end{center}
\end{figure*}

Finally, the mean SSIM and PSNR curves for selected 32-point transforms are shown in Fig.~\ref{F:MSSIM32} and Fig.~\ref{F:PSNR32}.
We exhibit the scaled DCT approximations $\mathbf{\tilde C}^{(1)}_{32}$, $\mathbf{\tilde C}^{(2)}_{32}$, $\mathbf{\tilde C}^{(6)}_{32}$, and $\mathbf{\tilde C}^{(7)}_{32}$, besides peering methods like BAS$_4$, OCBSML, and the exact DCT.
The transform
$\mathbf{\tilde C}^{(7)}_{32}$ performs better than all other competing approaches, except for OCBSML.
Although
$\mathbf{\tilde C}^{(1)}_{32}$
performs relatively poorly,
it saves up to 25\% of the total number of additions
when
compared with the other approaches and does not require any bit-shifting operation.
Fig.~\ref{F:LPIPS32}  shows that the DCT is outperformed by $\mathbf{\tilde C}^{(6)}_{32}$ and OCBSML for $r < 500$ values in terms of LPIPS.

Fig.~\ref{F:MSSIMadd32}, Fig.~\ref{F:PSNRadd32}, and ~\ref{F:LPIPSadd32} depict the SSIM, PSNR, and LPIPS gains per addition operation for the considered 32-point transforms, respectively.
The method $\mathbf{\tilde C}^{(1)}_{32}$ achieves the best SSIM and PSNR gains per addition unit when compared with the other approaches for practically any $r$ value.
The curves also show
that the proposed transform $\mathbf{\tilde C}^{(7)}_{32}$
presents consistently better results than the remaining approaches, $\mathbf{\tilde C}^{(6)}_{32}$, BAS$_4$, and OCBSML.
LPIPS by addition unit further highlights the results of $\mathbf{\tilde C}^{(6)}_{32}$ and OCBSML.

\begin{figure}[!h]
\begin{center}
\subfigure[PNSR = 31.6393 dB]{{\includegraphics[width=0.3\textwidth]{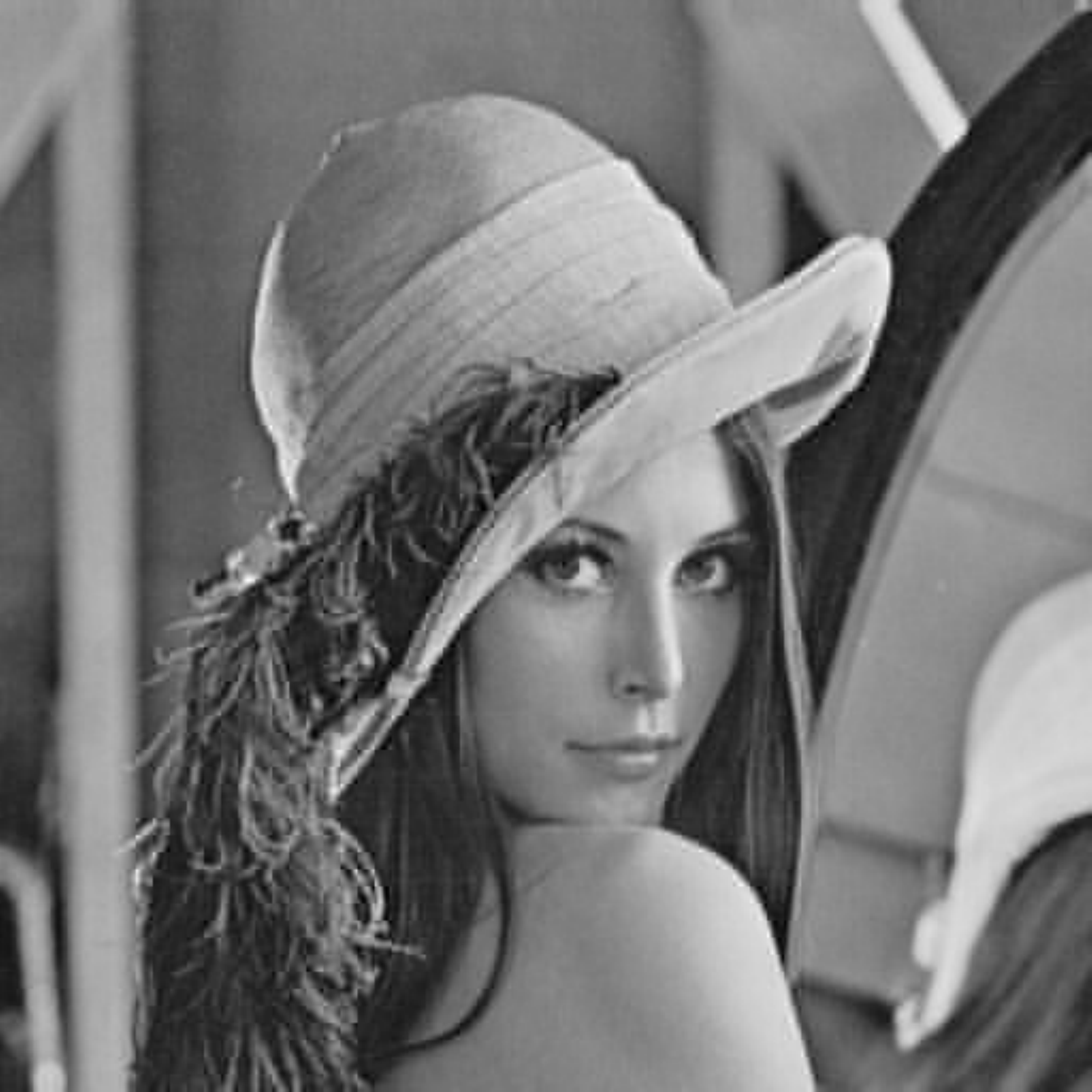}}}
\subfigure[PNSR = 26.6691 dB]{{\includegraphics[width=0.3\textwidth]{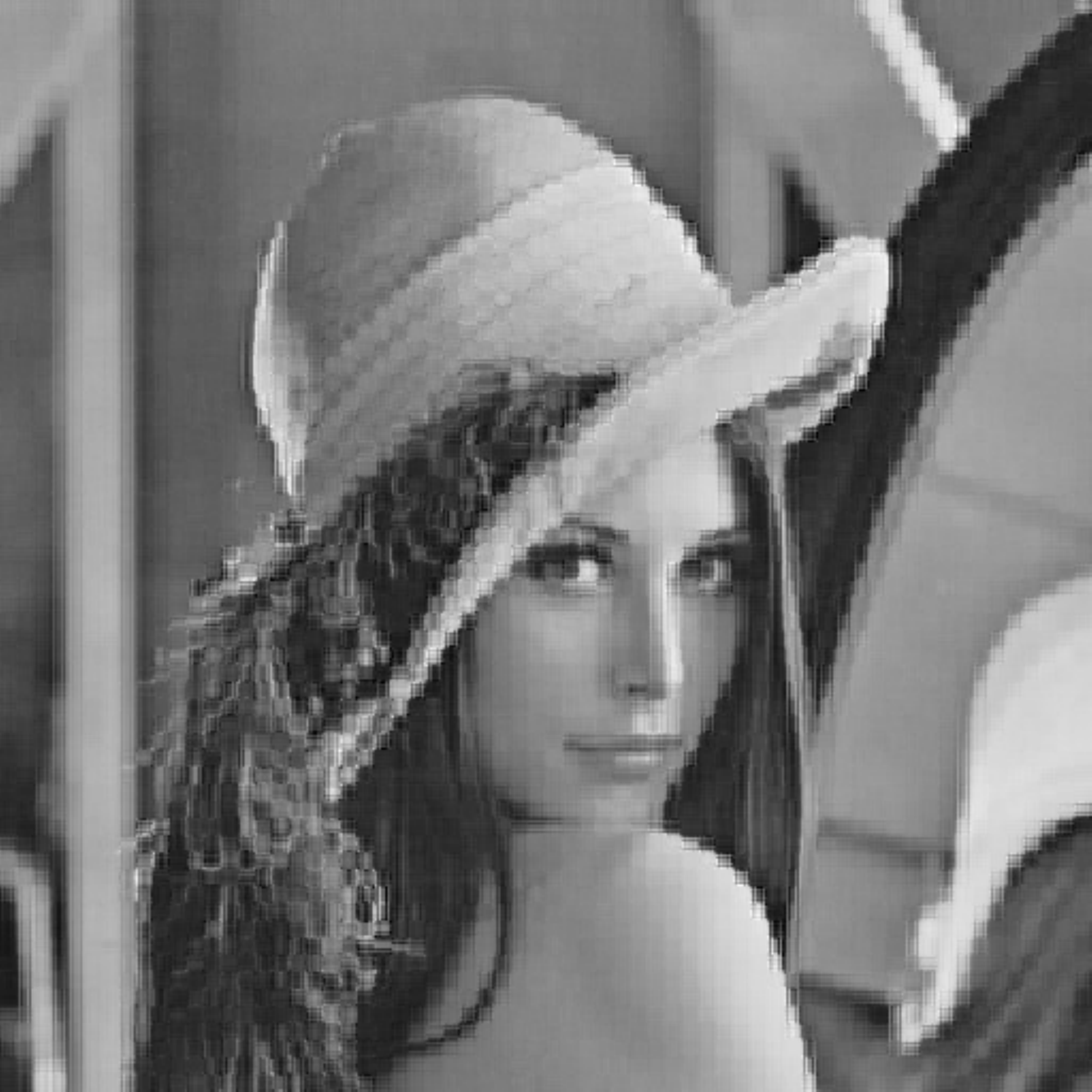}}}
\subfigure[PNSR = 30.4560 dB]{{\includegraphics[width=0.3\textwidth]{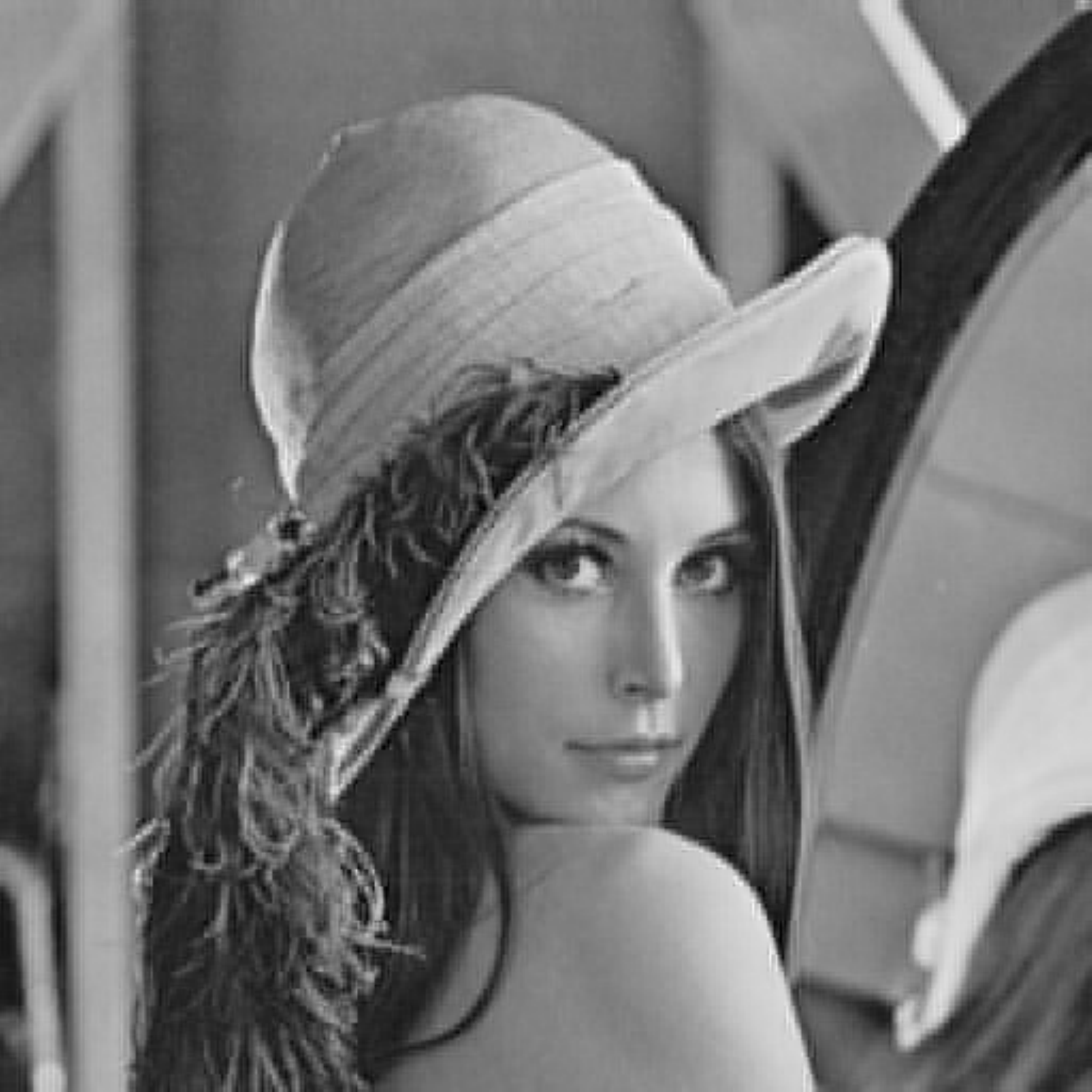}}}
\subfigure[PNSR = 32.0423 dB]{{\includegraphics[width=0.3\textwidth]{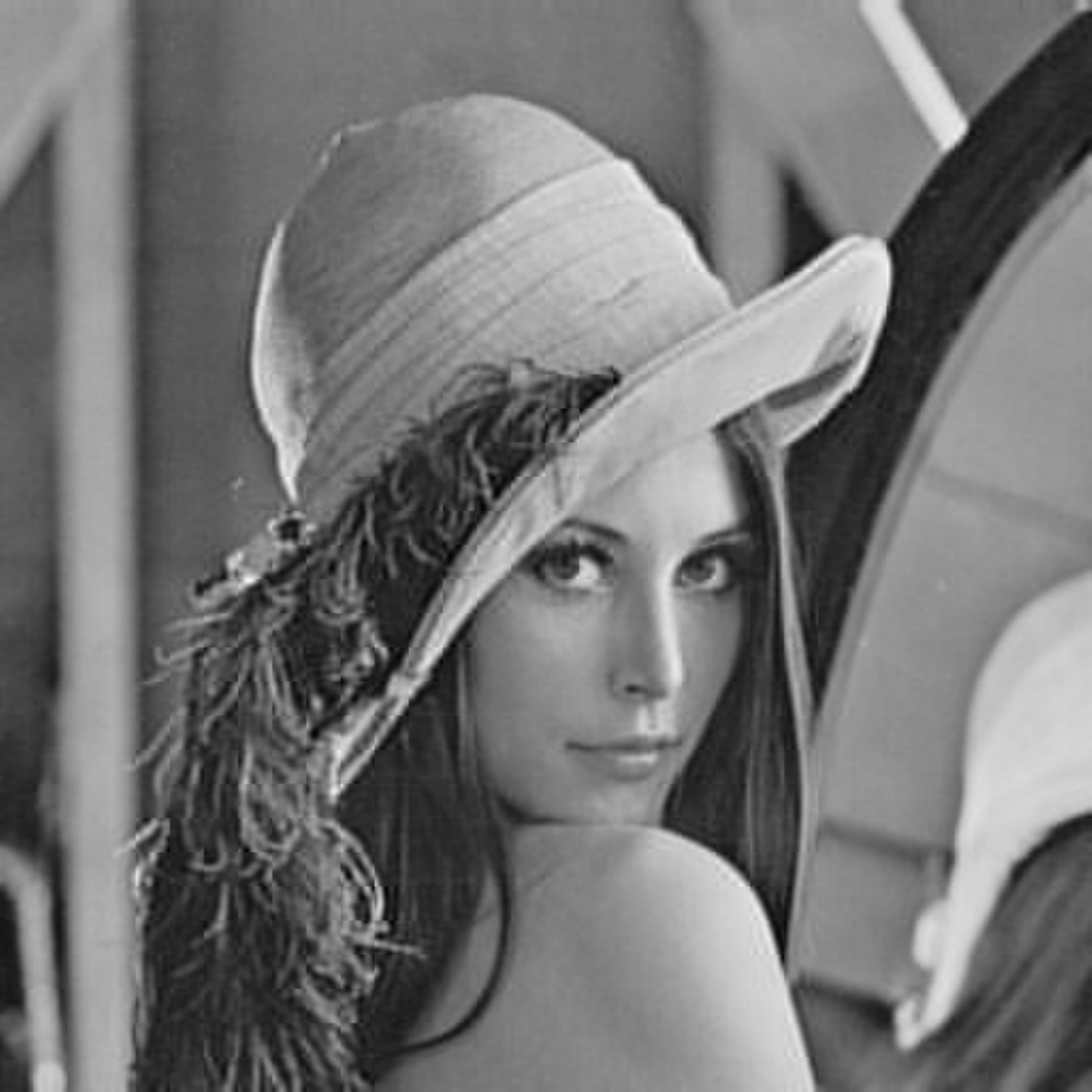}}}
\subfigure[PNSR = 26.6102 dB]{{\includegraphics[width=0.3\textwidth]{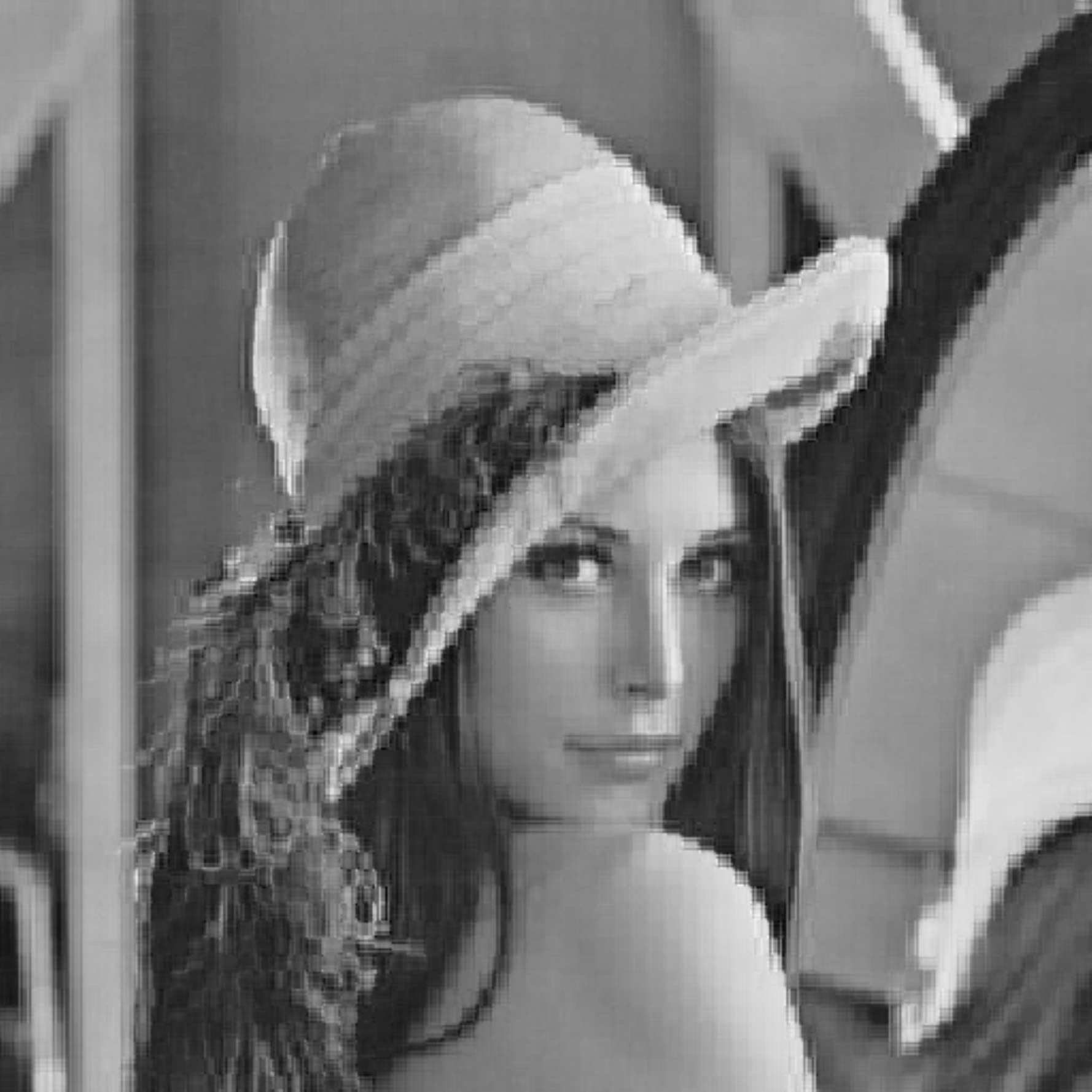}}}
\subfigure[PNSR = 30.3091 dB]{{\includegraphics[width=0.3\textwidth]{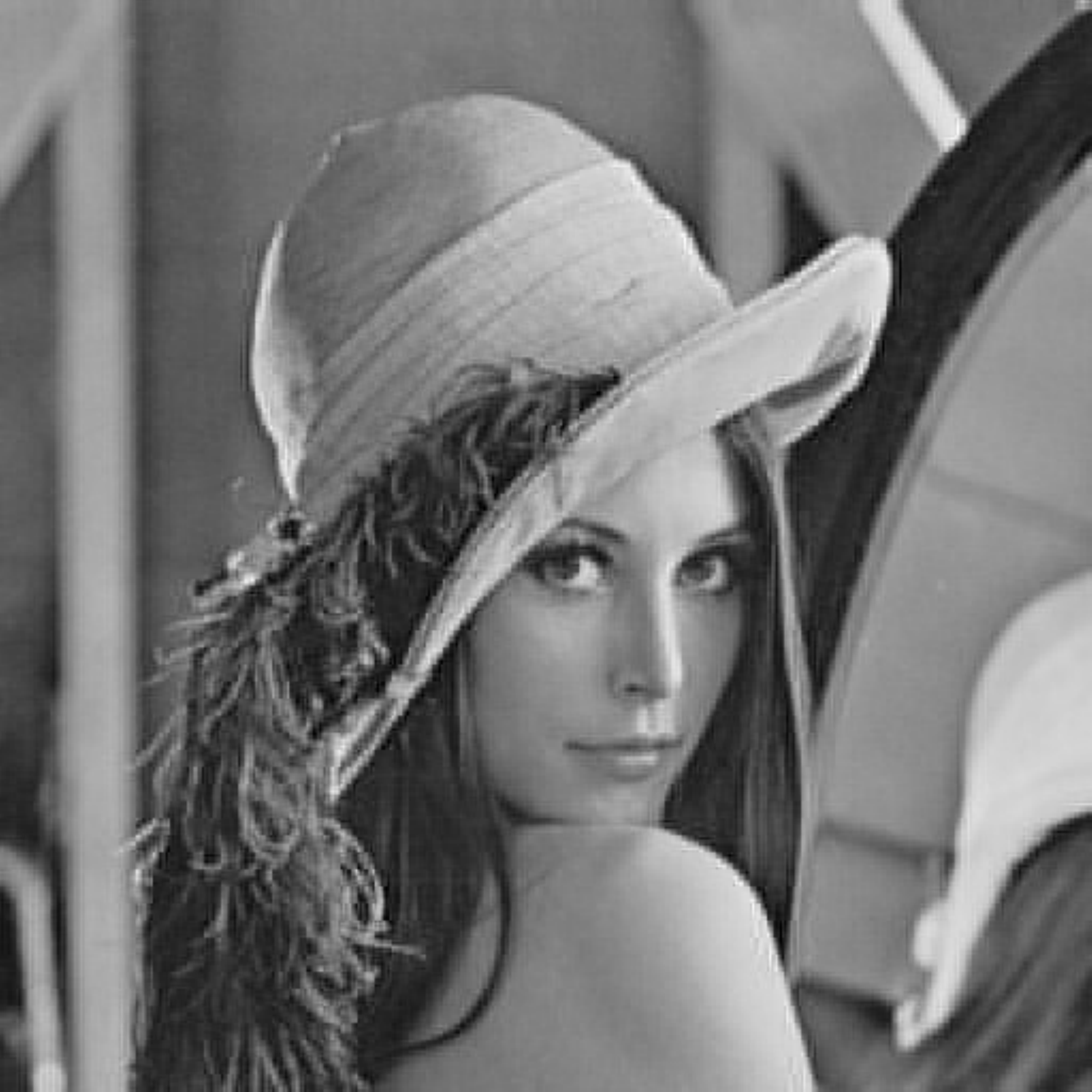}}}
\subfigure[PNSR = 32.5315 dB]{{\includegraphics[width=0.3\textwidth]{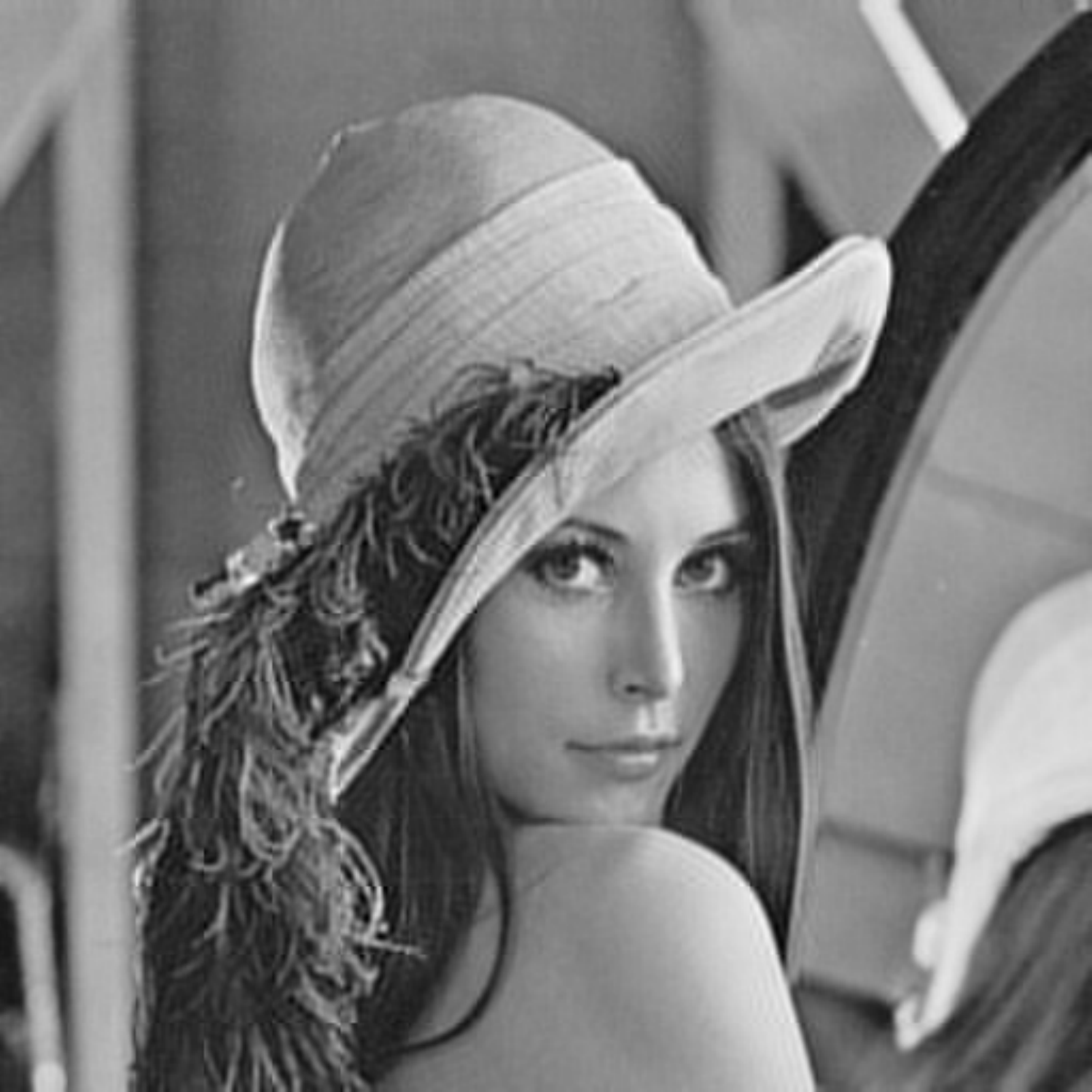}}}
\subfigure[PNSR = 26.6179 dB]{{\includegraphics[width=0.3\textwidth]{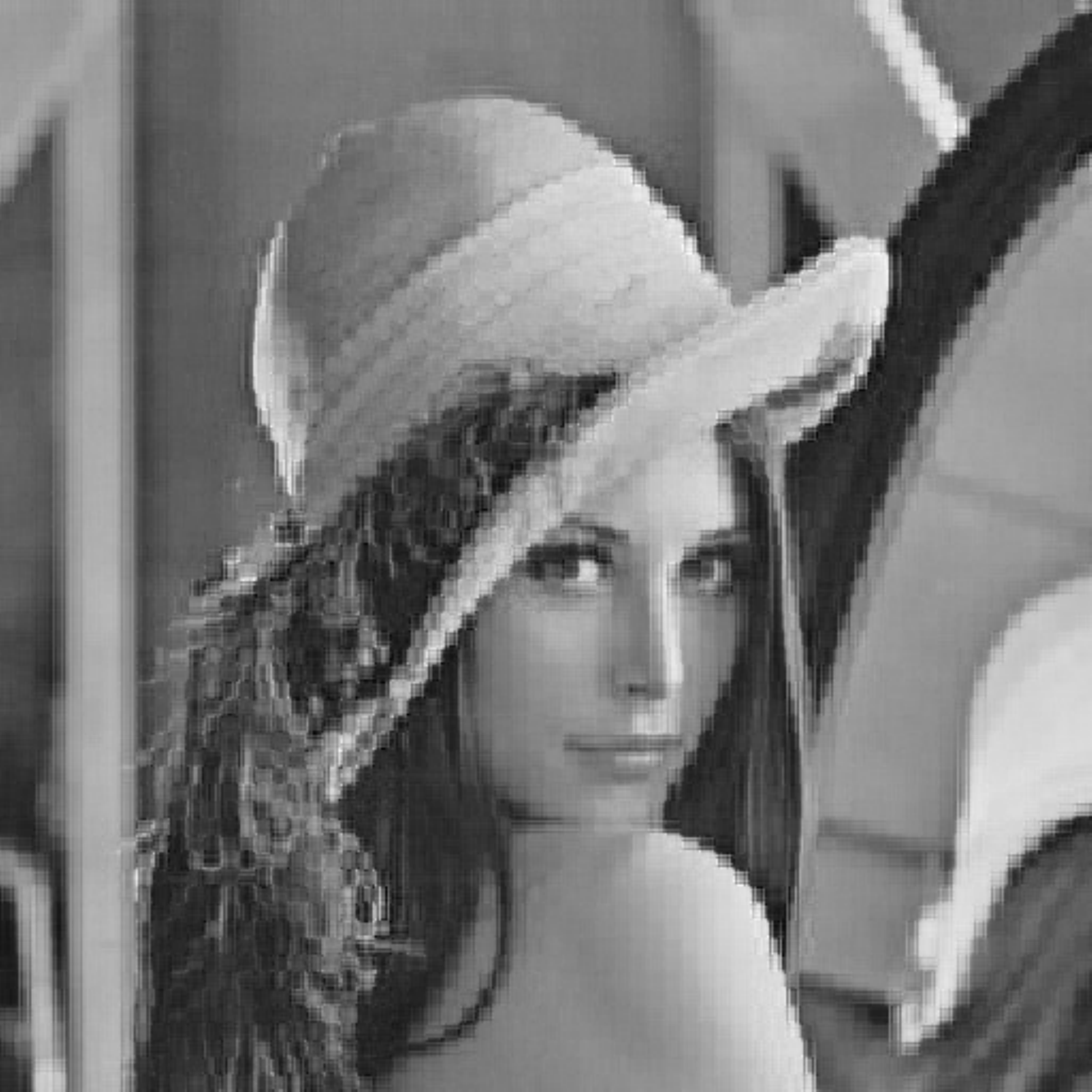}}}
\subfigure[PNSR = 30.3445 dB]{{\includegraphics[width=0.3\textwidth]{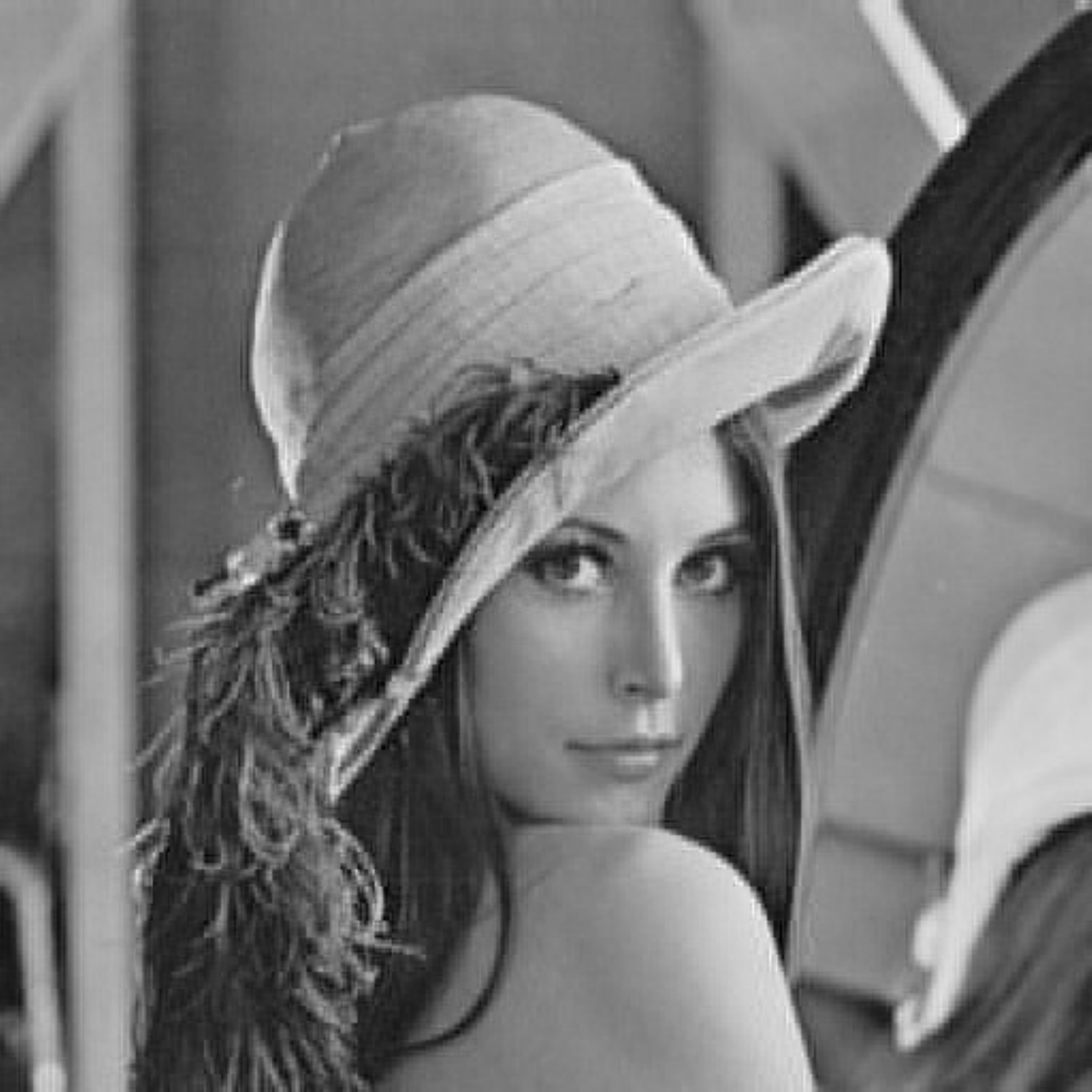}}}
\caption{``Lena'' image compressed by  (a)(d)(g) the DCT,
(b)(e)(h) $\mathbf{\tilde C}^{(1)}_{N}$, and (c)(f)(i) $\mathbf{\tilde C}^{(7)}_{N}$ for (a--c) $N = 8$, (d--f) $N = 16$, and (g--i) $N = 32$. The parameter
$r$ is set to 10, 40, and 155, for $N = 8, 16, 32$, respectively.}
\label{f:lena}
\end{center}
\end{figure}

Fig.~\ref{f:lena} exemplifies the compression of the ``Lena'' image by selected transforms: $\mathbf{\tilde C}^{(1)}_{N}$ and $\mathbf{\tilde C}^{(7)}_{N}$, and the  DCT for $N \in \{8, 16, 32\}$.
We select $r$ values so that the compression rate is roughly 84\%.
Note that transforms $\mathbf{\tilde C}^{(7)}_{8}$, $\mathbf{\tilde C}^{(7)}_{16}$, and $\mathbf{\tilde C}^{(7)}_{32}$ have high PSNR values and virtually no visual degradation.

\subsection{Video Coding}

This section reports the suitability results
of the introduced best-performing 8-point transform---$\mathbf{\tilde{C}}_{8}^{(7)}$---together with its scaled 16- and 32-point versions to video coding.
We also included in our analysis the optimal 8-point transform $\mathbf{\tilde{C}}_{8}^{(1)}$ (MRDCT) and its scaled versions of lengths 16 and 32 since they possess a very low arithmetic complexity and still compelling coding results.
In this experiment, we embedded the two groups of 8-, 16- and 32-point transforms into a publicly available HEVC reference software~\cite{refsoft}, and then assessed the performance of the resulting systems.
For simplicity, hereafter we refer to these groups as $\mathbf{\tilde{C}}^{(1)}_{\text{group}}$ and $\mathbf{\tilde{C}}^{(7)}_{\text{group}}$.
The core HEVC integer DCT (intDCT) transforms of lengths 8, 16 and 32 require 50, 186, and 682 additions and 30, 86, and 287 bit-shifts, respectively~\cite{Meher2014}.
The 4-point intDCT HEVC transform requires no approximation
because it is already a low-complexity
multiplierless
transformation.

We encoded the first 100~frames of representative video sequences
from each A to F class
according to
the recommendations
detailed in the Common Test Conditions (CTC) document~\cite{CTConditions2013}.
The following 8-bit videos were selected:
``PeopleOnStreet'' (2560$\times$1600 at 30~fps),
``BasketballDrive'' (1920$\times$1080 at 50~fps),
``RaceHorses'' (832$\times$480 at 30~fps),
``BlowingBubbles'' (416$\times$240  at 50~fps),
``KristenAndSara'' (1280$\times$720  at 60~fps),
and
``BasketballDrillText'' (832$\times$480  at 50~fps).
We set the encoding parameters for the Main profile and All-Intra (AI), Random Access (RA), Low Delay B (LD-B), and Low Delay P (LD-P) configurations according to the CTC document.
Our experiments consider varying the quantization parameter (QP) in $\{22, 27, 32, 37\}$ as recommended by the CTC documentation~\cite{CTConditions2013}.

\begin{figure}[!h]
\begin{center}
\subfigure[AI]{{\includegraphics[height=0.3\textwidth]{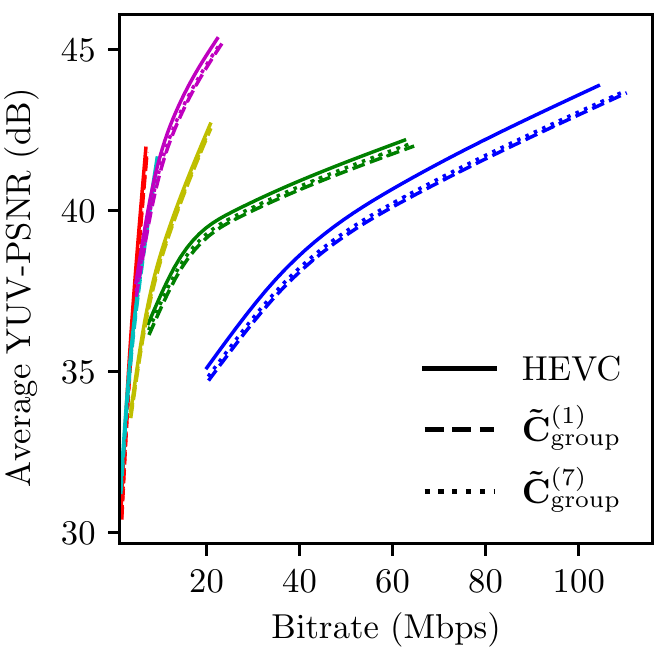}}}
\qquad
\subfigure[RA]{{\includegraphics[height=0.3\textwidth]{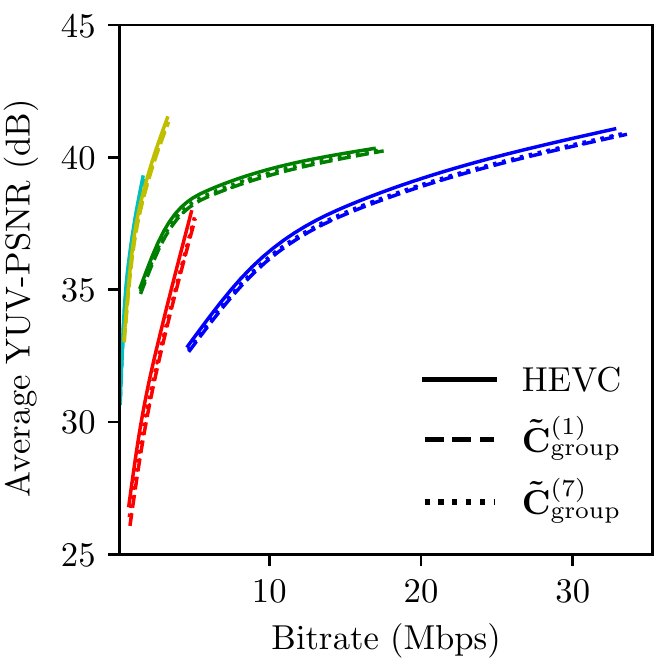}}}

\subfigure[LD-B]{{\includegraphics[height=0.3\textwidth]{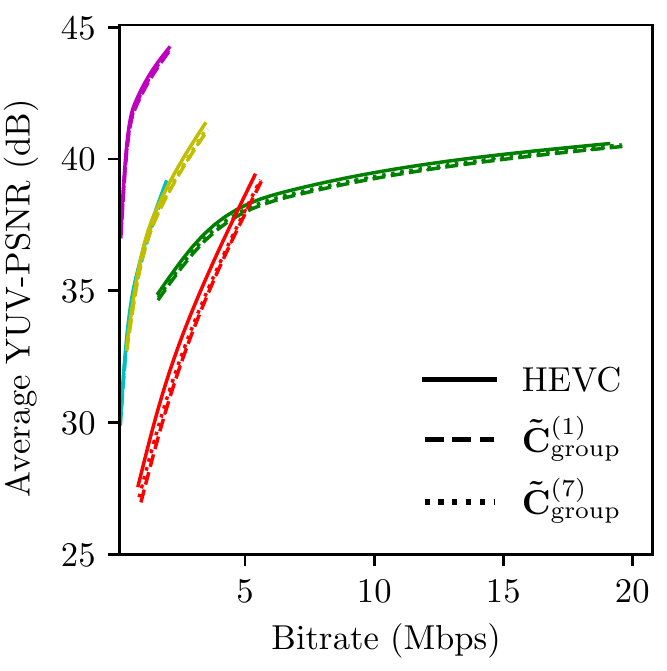}}}
\qquad
\subfigure[LD-P]{{\includegraphics[height=0.3\textwidth]{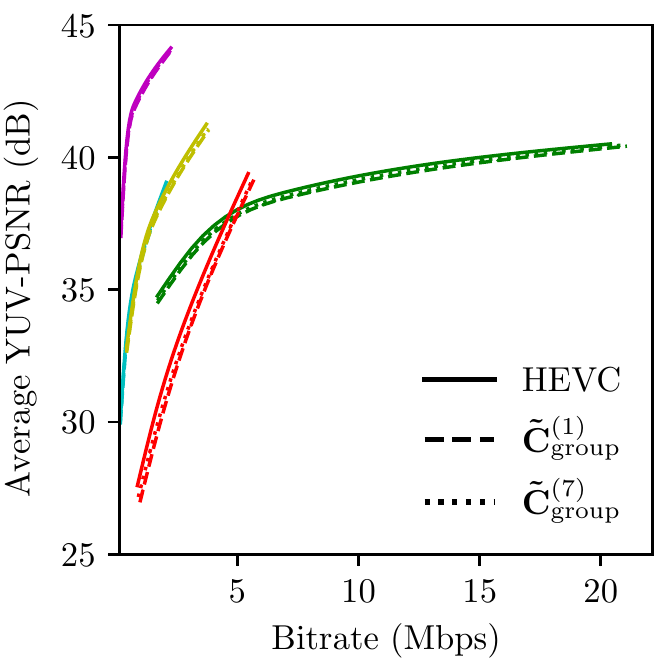}}}
\caption{RD curves for different configuration modes and video sequences. Curves associated to ``PeopleOnStreet'', ``BasketballDrive'', ``RaceHorses'', ``BlowingBubbles'', ``BasketballDrillText'', and ``KristenAndSara'' are represented by colors blue, green, red, cyan, yellow, and magenta, respectively.}
\label{F:rdcurves3}
\end{center}
\end{figure}

The reference software itself measures the MSE and PSNR for each frame and color channel (in YUV color space), and we compute the YUV-PSNR~\cite{Ohm2012}.
We used these measurements for each QP value
for computing the Bj{\o}ntegaard's delta PSNR (BD-PSNR) and delta rate (BD-Rate)~\cite{Bjontegaard2001, Hanhart2014} for the two groups of transforms and the four coding configurations.
Table~\ref{tab:bdpsnrbdrate} lists the obtained BD-PSNR and BD-Rate measurements.
Fig.~\ref{F:rdcurves3} presents the corresponding rate-distortion (RD) curves, which are interpolated by cubic splines for better visualization~\cite{Coelho2018,Hanhart2014}.
Although the proposed transforms present some quality loss, they require a fraction of the required operations by HEVC core transforms.

The transform $\mathbf{\tilde{C}}^{(7)}_{\text{group}}$
obtained much better results than $\mathbf{\tilde{C}}^{(1)}_{\text{group}}$ for all configuration modes and video sequences.
These findings corroborate the results from the still-image compression experiments from Section~\ref{subsec:image}.
Note that replacing the original HEVC transforms by $\mathbf{\tilde{C}}^{(7)}_{\text{group}}$ results in a loss of no more than 0.54~dB in AI configuration, which may be negligible depending on the  target application.

\begin{table}[!h]
\centering
\setlength{\tabcolsep}{2pt}
\caption{Average BD-PSNR and BD-Rate for the transforms  $\mathbf{\tilde{C}}^{(1)}_{\text{group}}$ and $\mathbf{\tilde{C}}^{(7)}_{\text{group}}$ embedded into the HEVC reference software}
\label{tab:bdpsnrbdrate}
\begin{tabular}{cccccccc}
\hline
\multirow{2}{*}{ Config. } & \multirow{2}{*}{ Video sequence } & \multicolumn{2}{c}{ BD-PSNR (dB) } & \multicolumn{2}{c}{ BD-Rate (\%) } \\
\cline{3-6}
& & $\mathbf{\tilde{C}}^{(1)}_{\text{group}}$ & $\mathbf{\tilde{C}}^{(7)}_{\text{group}}$ & $\mathbf{\tilde{C}}^{(1)}_{\text{group}}$ & $\mathbf{\tilde{C}}^{(7)}_{\text{group}}$  \\
\hline
\multirow{6}{*}{ AI }
& ``PeopleOnStreet'' & $-0.5405$ &  $\mathbf{-0.4156}$ &  $10.7981$  &  $\mathbf{8.2472}$  \\
& ``BasketballDrive'' & $-0.3312$ &  $\mathbf{-0.2013}$ &  $13.0744$  &  $\mathbf{7.7961}$  \\
& ``RaceHorses'' & $-0.6681$ &  $\mathbf{-0.5373}$ &  $8.7455$  &  $\mathbf{7.0180}$  \\
& ``BlowingBubbles'' & $-0.2569$ &  $\mathbf{-0.1568}$ &  $4.5457$  &  $\mathbf{2.7863}$  \\
& ``KristenAndSara'' & $-0.4717$ &  $\mathbf{-0.3225}$ &  $9.8382$  &  $\mathbf{6.6716}$  \\
& ``BasketballDrillText'' & $-0.2049$ &  $\mathbf{-0.1320}$ &  $4.0161$  &  $\mathbf{2.5766}$  \\
\hline
\multirow{5}{*}{ RA }
& ``PeopleOnStreet'' & $-0.2908$ &  $\mathbf{-0.2132}$ &  $7.1735$  &  $\mathbf{5.2135}$  \\
& ``BasketballDrive'' & $-0.2560$ &  $\mathbf{-0.1609}$ &  $12.1530$  &  $\mathbf{7.5263}$  \\
& ``RaceHorses'' & $-0.9507$ &  $\mathbf{-0.6369}$ &  $16.0780$  &  $\mathbf{10.9197}$  \\
& ``BlowingBubbles'' & $-0.2009$ &  $\mathbf{-0.1119}$ &  $5.4653$  &  $\mathbf{3.0192}$  \\
& ``BasketballDrillText'' & $-0.2490$ &  $\mathbf{-0.1704}$ &  $6.2465$  &  $\mathbf{4.2235}$  \\
\hline
\multirow{5}{*}{ LD-B }
& ``BasketballDrive'' & $-0.2381$ &  $\mathbf{-0.1537}$ &  $10.6718$  &  $\mathbf{6.7131}$  \\
& ``RaceHorses'' & $-0.9274$ &  $\mathbf{-0.6399}$ &  $14.5320$  &  $\mathbf{10.1138}$  \\
& ``BlowingBubbles'' & $-0.2023$ &  $\mathbf{-0.1122}$ &  $5.6422$  &  $\mathbf{3.0933}$  \\
& ``KristenAndSara'' & $-0.2351$ &  $\mathbf{-0.1640}$ &  $8.2957$  &  $\mathbf{5.7539}$  \\
& ``BasketballDrillText'' & $-0.2895$ &  $\mathbf{-0.2054}$ &  $7.6528$  &  $\mathbf{5.3746}$  \\
\hline
\multirow{5}{*}{ LD-P }
& ``BasketballDrive'' & $-0.2411$ &  $\mathbf{-0.1539}$ &  $10.6911$  &  $\mathbf{6.7416}$  \\
& ``RaceHorses'' & $-0.8873$ &  $\mathbf{-0.6207}$ &  $13.7702$  &  $\mathbf{9.7240}$  \\
& ``BlowingBubbles'' & $-0.1890$ &  $\mathbf{-0.1087}$ &  $5.3818$  &  $\mathbf{3.0557}$  \\
& ``KristenAndSara'' & $-0.2227$ &  $\mathbf{-0.1493}$ &  $8.1583$  &  $\mathbf{5.5457}$  \\
& ``BasketballDrillText'' & $-0.2697$ &  $\mathbf{-0.1960}$ &  $7.2204$  &  $\mathbf{5.2078}$  \\
\hline
\end{tabular}
\end{table}

\begin{figure*}[!h]
\centering
\subfigure[
YUV-PSNR = 42.5687 dB
]{\includegraphics[width=.3\linewidth]{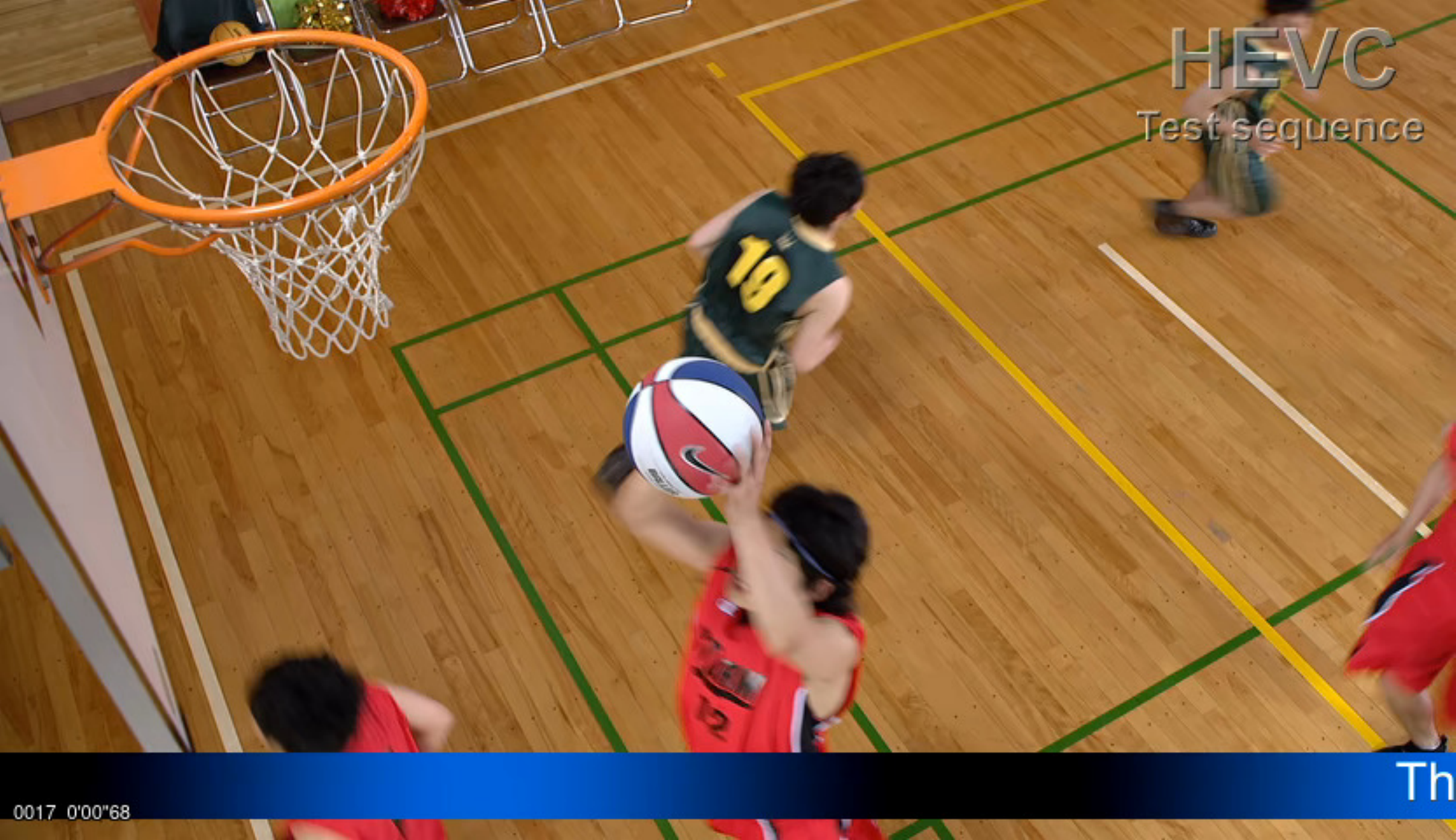}}\,
\subfigure[
YUV-PSNR = 37.7729 dB
]{\includegraphics[width=.3\linewidth]{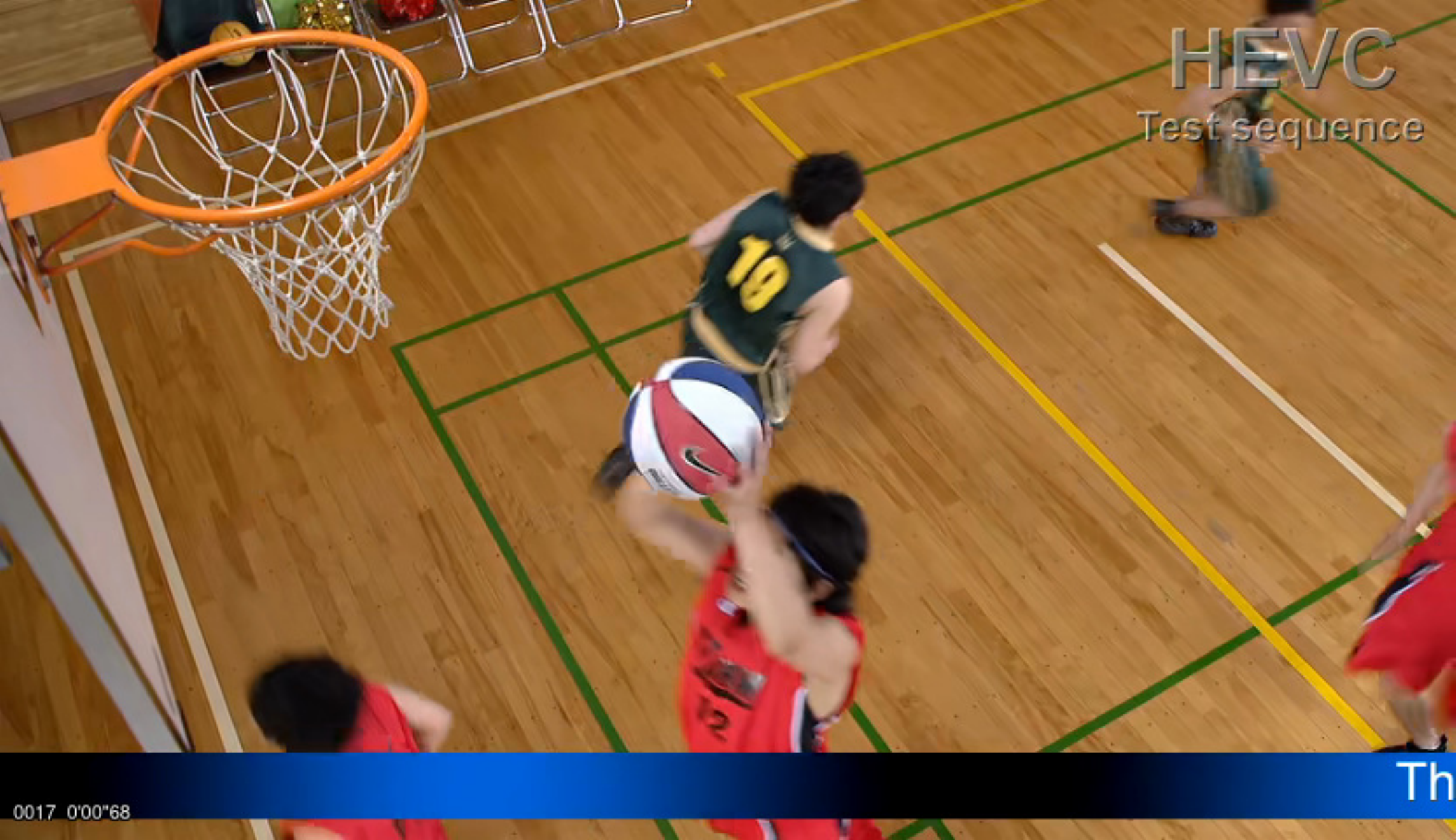}}\,
\subfigure[
YUV-PSNR = 35.1120 dB]{\includegraphics[width=.3\linewidth]{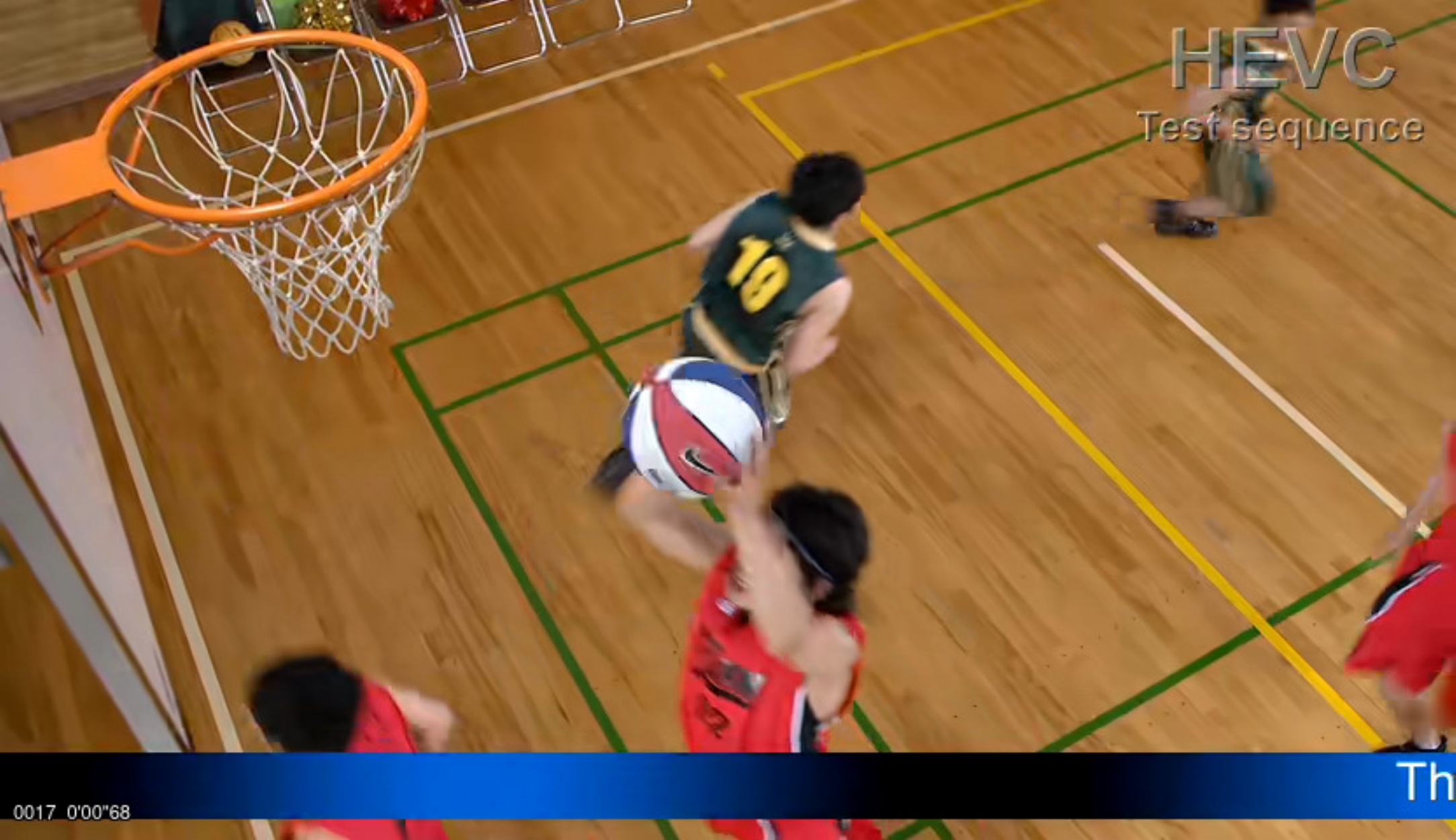}}

\subfigure[
YUV-PSNR = 32.6642 dB]{\includegraphics[width=.3\linewidth]{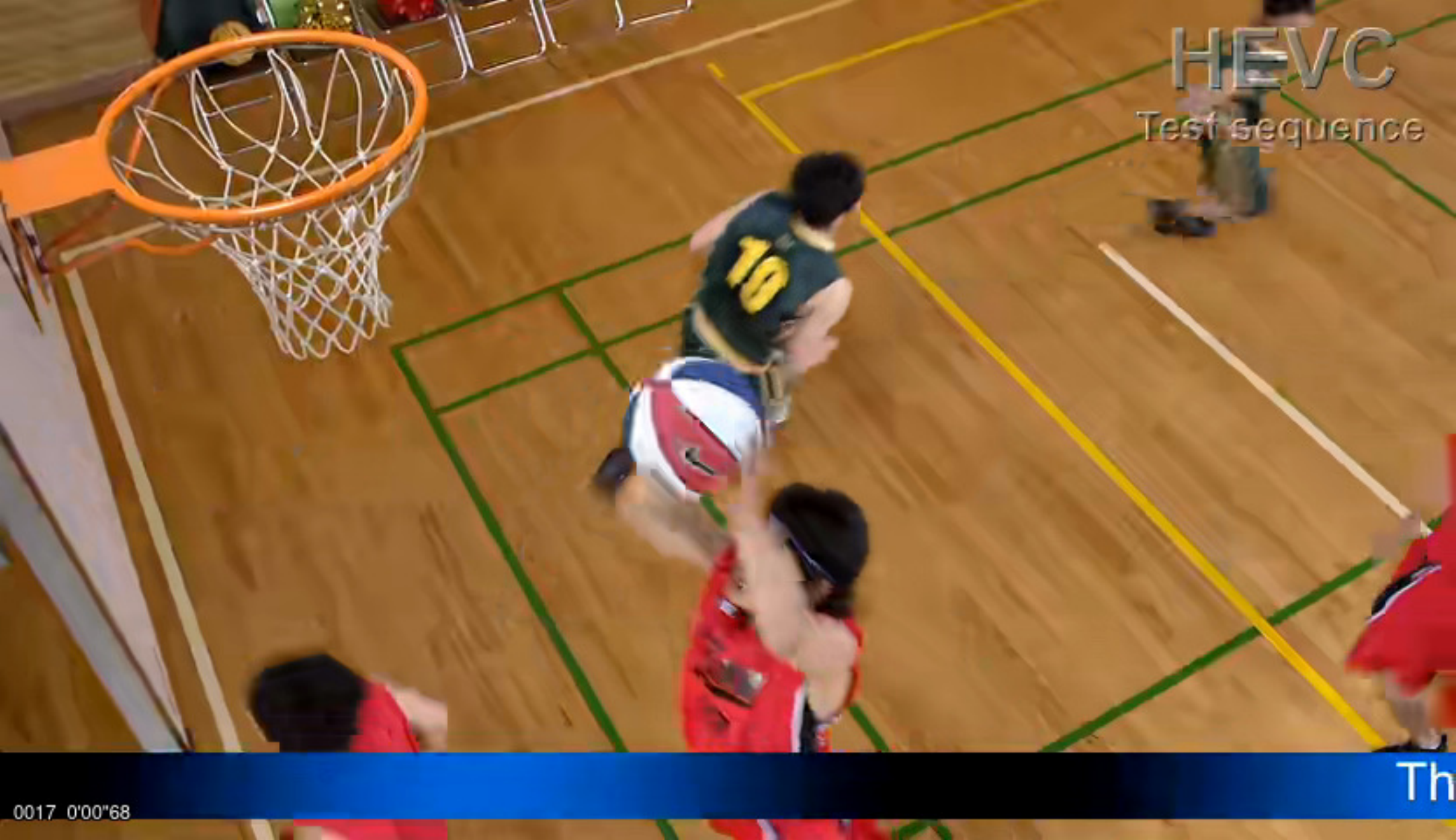}}\,
\subfigure[
YUV-PSNR = 42.4473 dB]{\includegraphics[width=.3\linewidth]{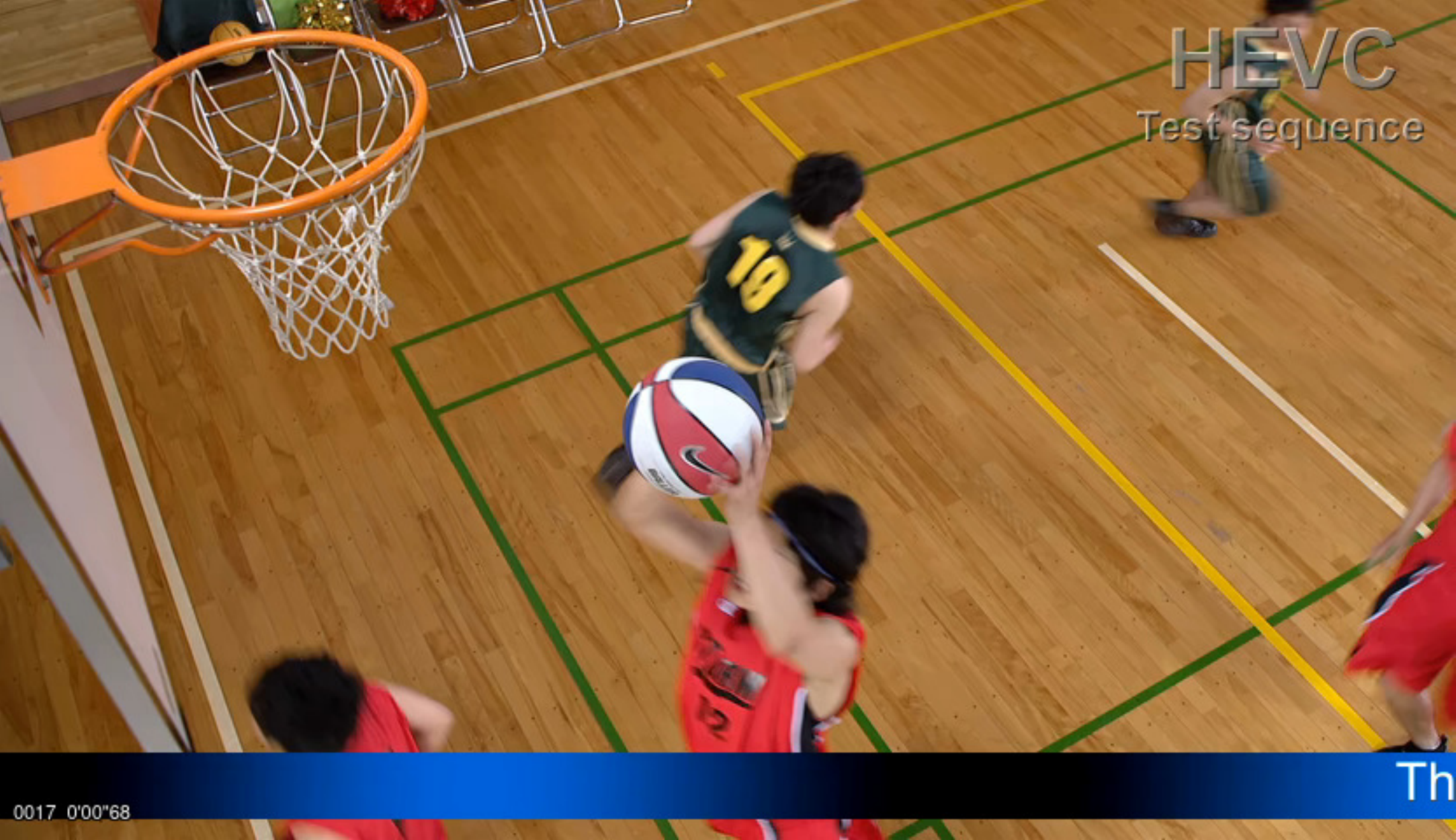}}\,
\subfigure[
YUV-PSNR = 37.6475 dB]{\includegraphics[width=.3\linewidth]{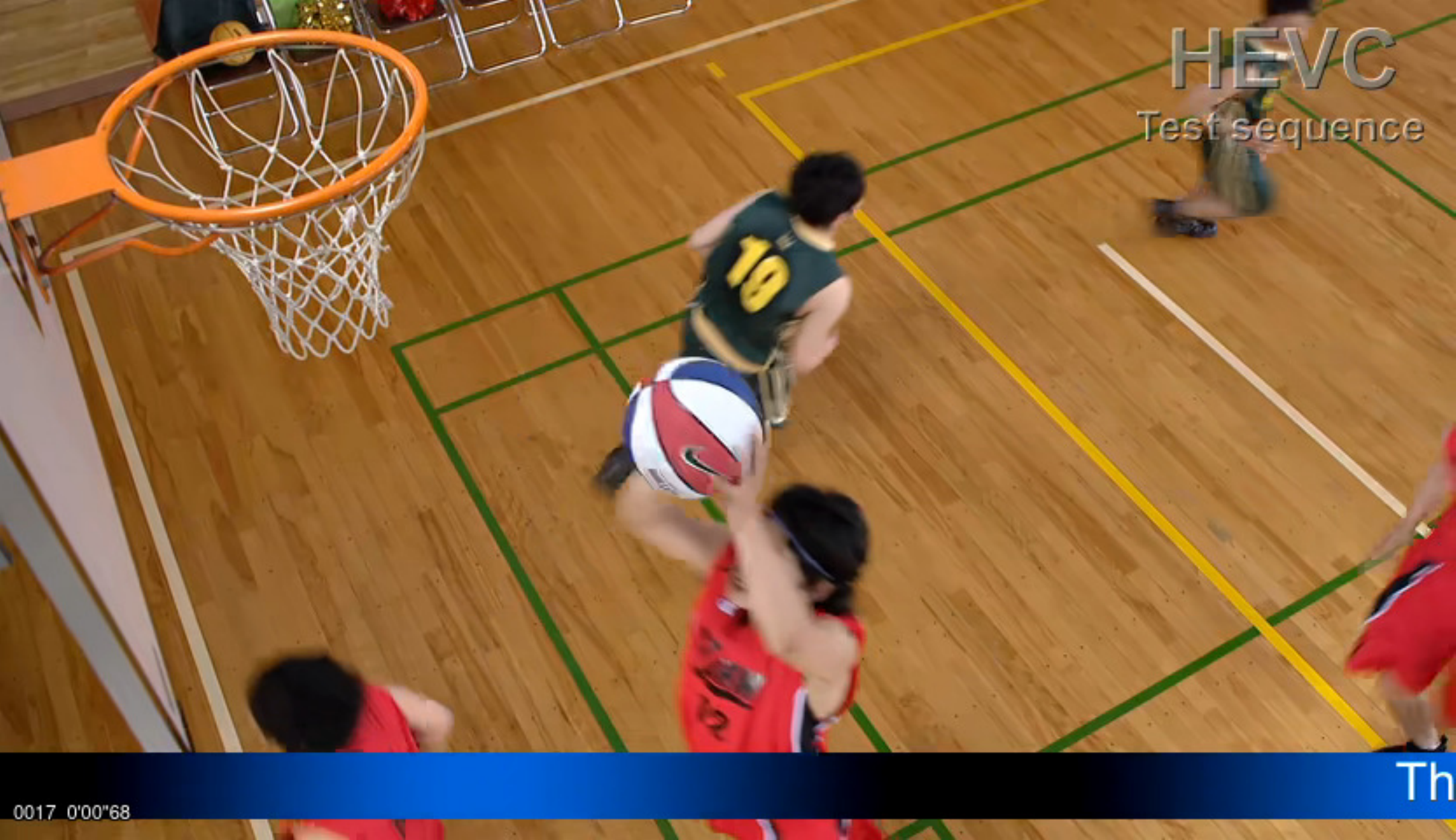}}

\subfigure[
YUV-PSNR = 34.9257 dB]{\includegraphics[width=.3\linewidth]{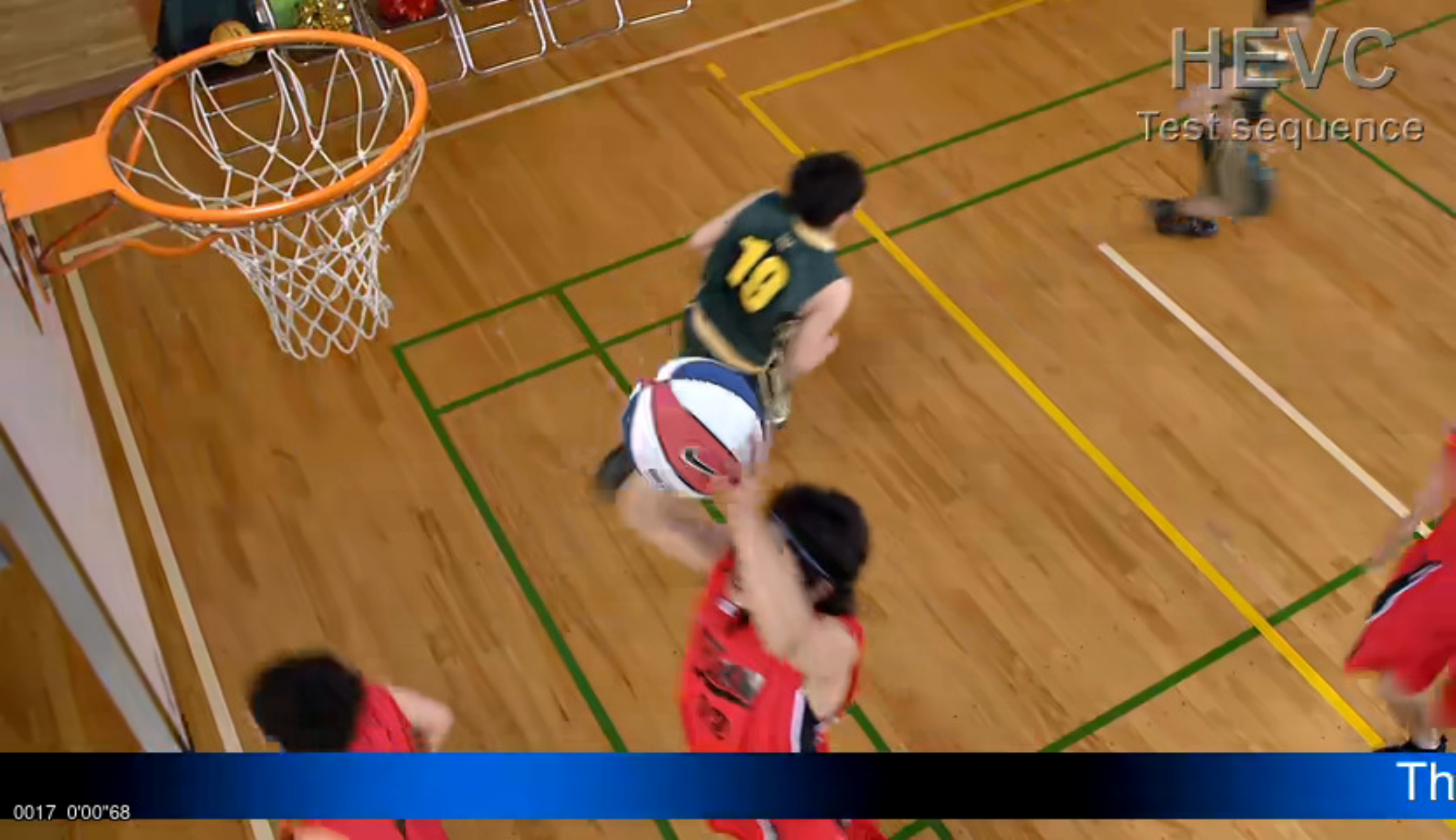}}\,
\subfigure[
YUV-PSNR = 32.4855 dB]{\includegraphics[width=.3\linewidth]{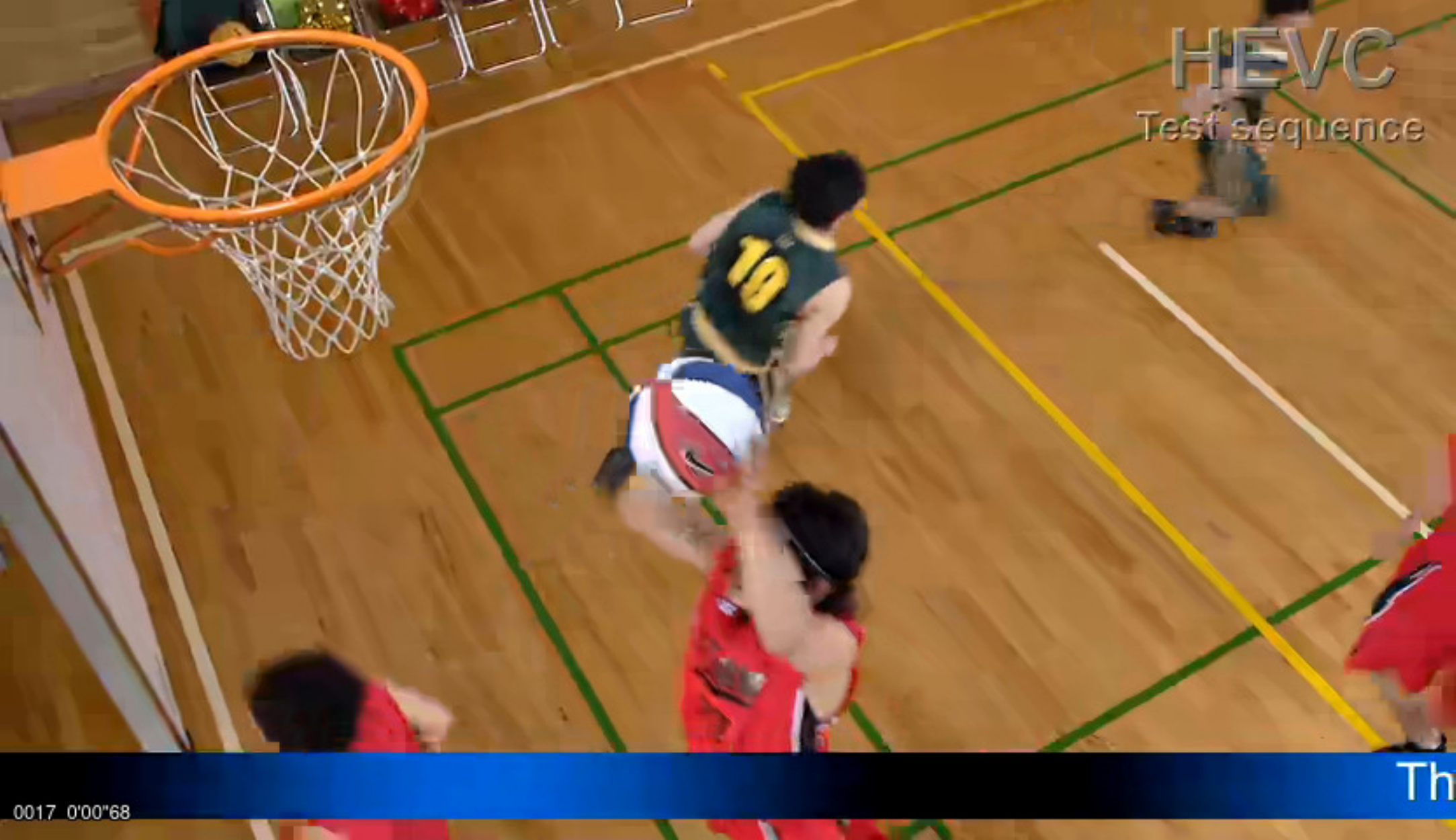}}\,
\subfigure[
YUV-PSNR = 42.4770 dB]{\includegraphics[width=.3\linewidth]{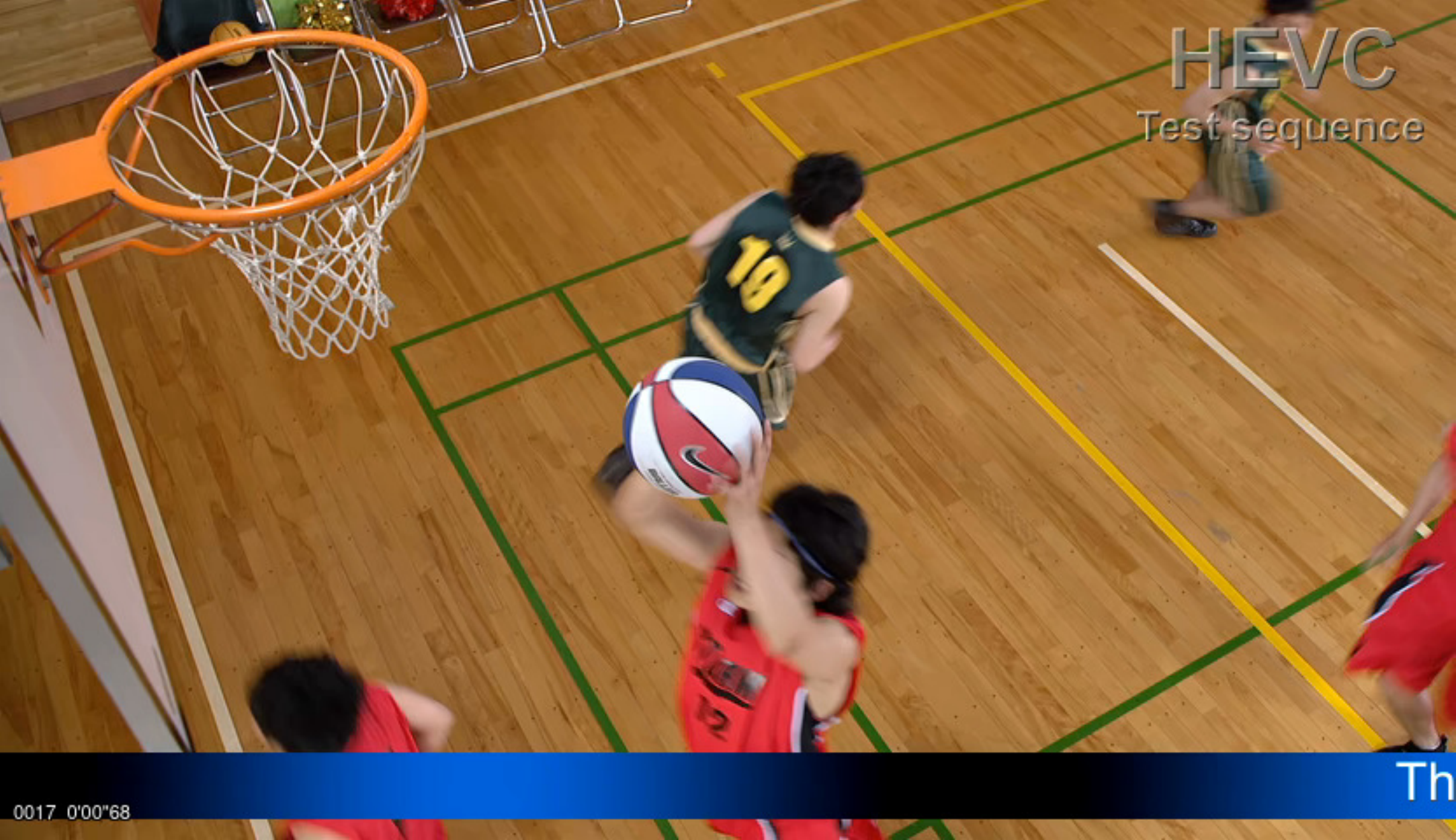}}

\subfigure[
YUV-PSNR = 37.6561 dB]{\includegraphics[width=.3\linewidth]{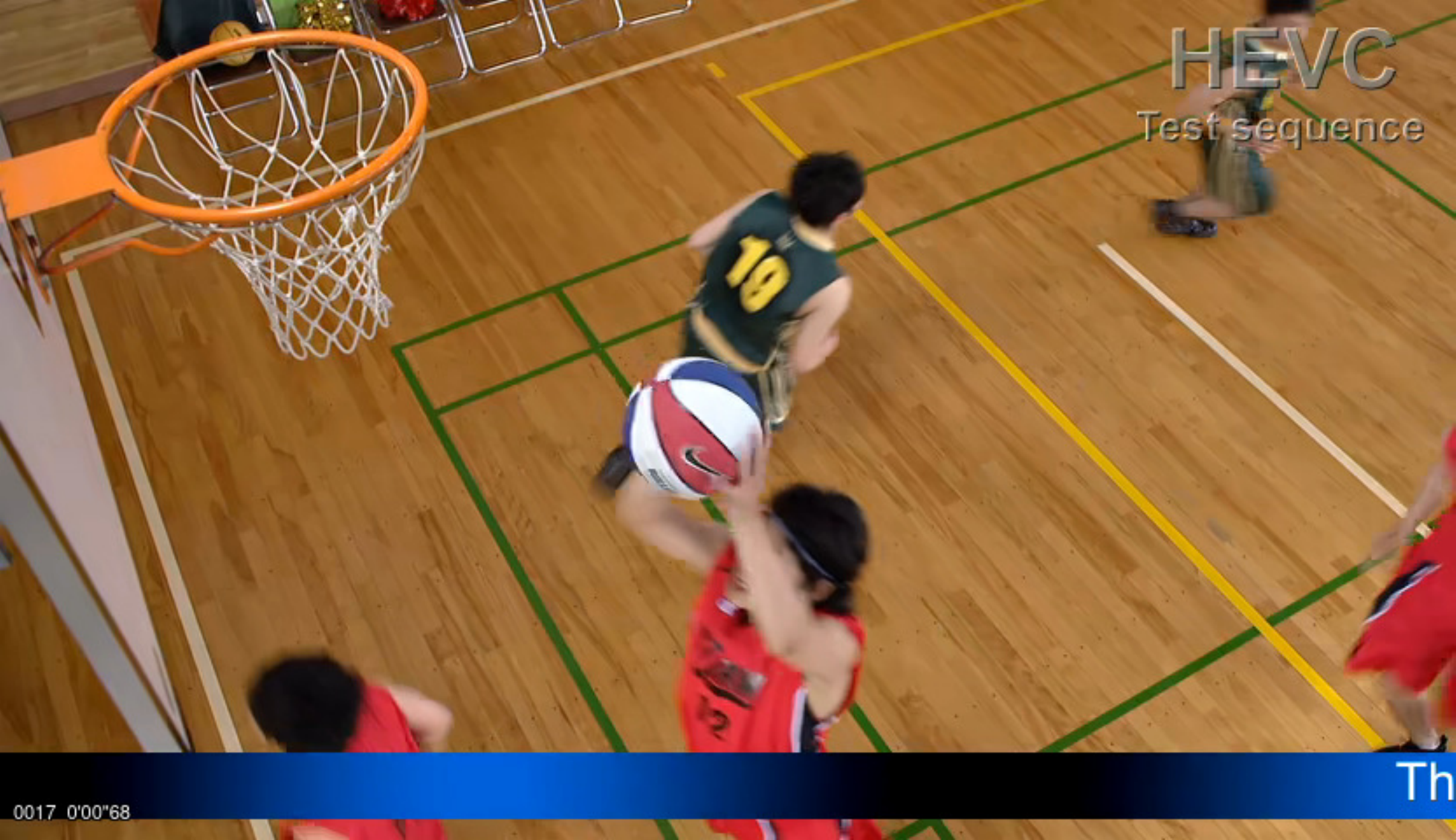}}\,
\subfigure[
YUV-PSNR = 35.0318 dB]{\includegraphics[width=.3\linewidth]{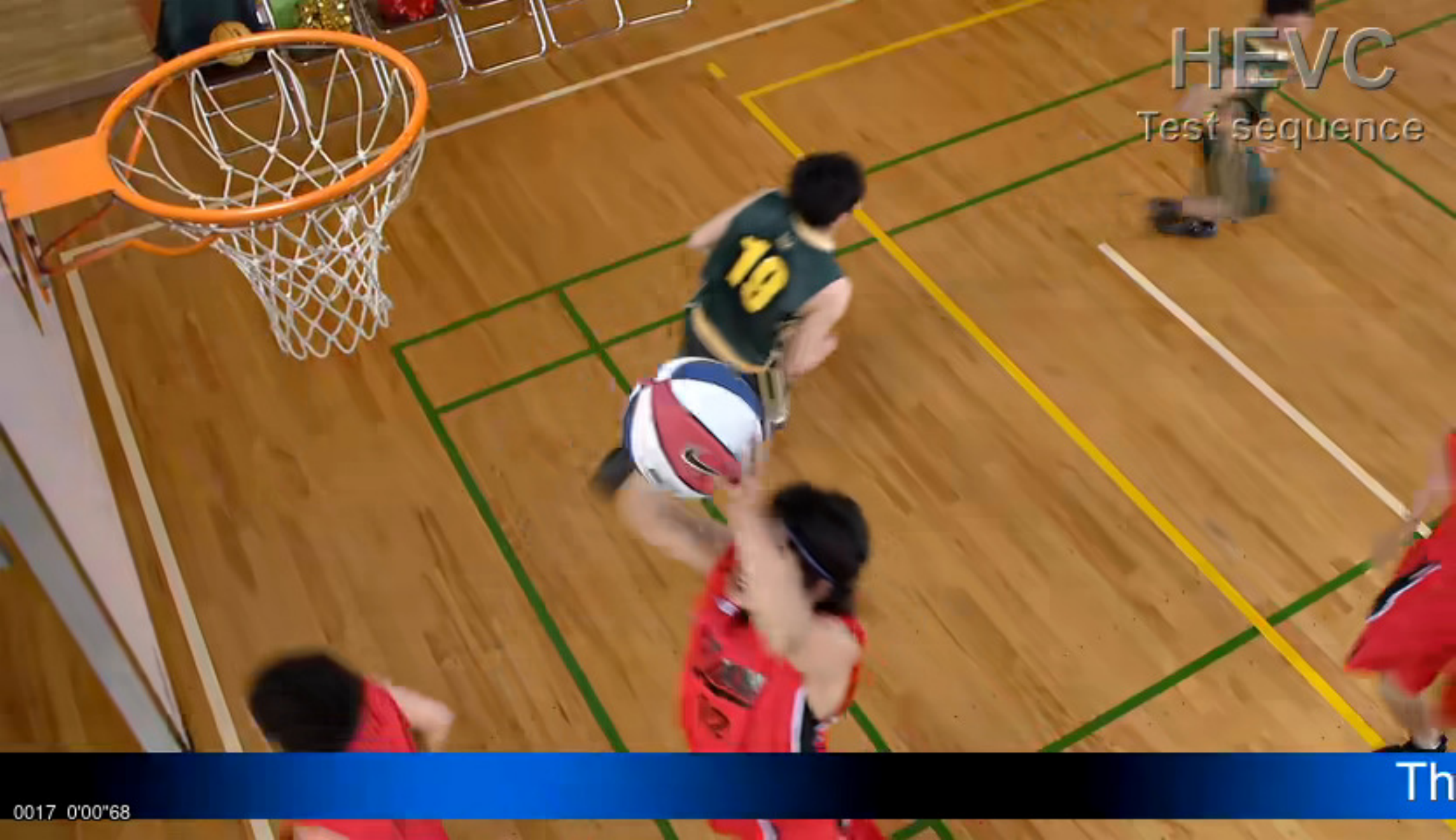}}\,
\subfigure[
YUV-PSNR = 32.5330 dB]{\includegraphics[width=.3\linewidth]{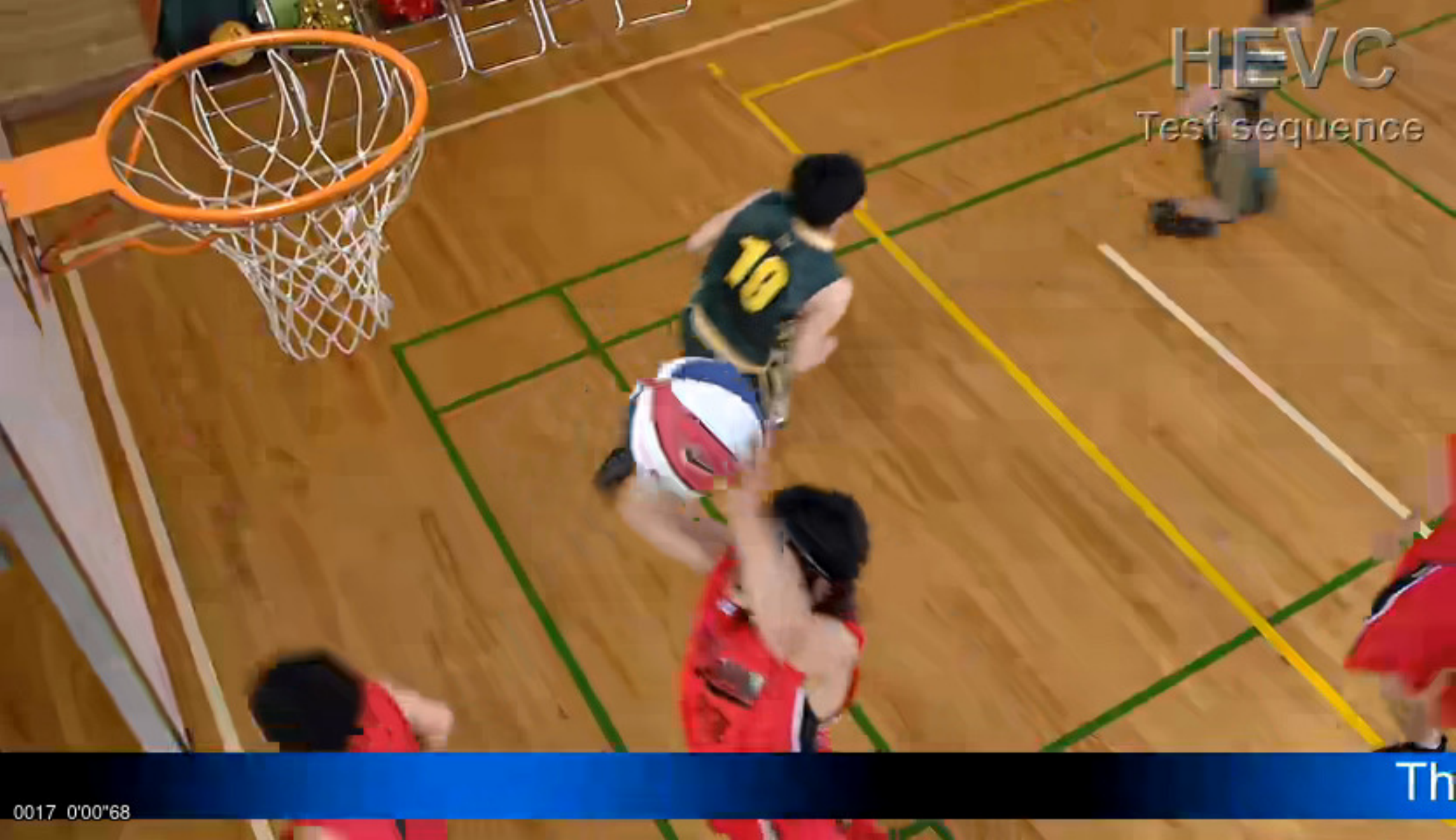}}
\caption{
Compressed frame of ``BasketballDrillText'' using the (a--d) original  HEVC, (e--h) $\mathbf{\tilde{C}}^{(1)}_{\text{group}}$ and (i--l) $\mathbf{\tilde{C}}^{(7)}_{\text{group}}$ transforms under different configuration modes and QP values.
Results for (a)(e)(i) AI and $\mbox{QP} = 22$, (b)(f)(j) RA and $\mbox{QP} = 27$, (c)(g)(k) LD-B and $\mbox{QP} = 32$, and (d)(h)(l) LD-P and $\mbox{QP} = 37$. }
\label{fig:exemplehevc}
\end{figure*}

Fig.~\ref{fig:exemplehevc} depicts the
eighteenth
frame of the ``BasketballDrillText'' video sequence encoded using the original HEVC transform suit
compared with
the results from
the transform groups
$\mathbf{\tilde{C}}^{(1)}_{\text{group}}$ and $\mathbf{\tilde{C}}^{(7)}_{\text{group}}$.
We also provide the YUV-PSNR measurements for the selected frame.
These results consider the
four
configuration modes and QPs.
Note that there is no visually noticeable degradation associated to the approximate DCTs.
These results suggest that the original HEVC transforms can be substituted
by
$\mathbf{\tilde{C}}^{(1)}_{\text{group}}$ or $\mathbf{\tilde{C}}^{(7)}_{\text{group}}$ without significant losses in image/video quality.

\section{Hardware Implementation} \label{sec:hardware}

The proposed 8-point low-complexity
transforms
were implemented on a field programmable gate array (FPGA).
The device adopted for the hardware
implementation was
the Xilinx Artix-7
XC7A35T-1CPG236C.

The designs use pipelined systolic architecture
for implementing each of the transforms~\cite{Baghaie2000, Safiri1996}.
The implemented blocks compute the transform using
the fast algorithm outlined in \eqref{eq:fastalg} and displayed in Fig.~\ref{fig:sfg}.
Each matrix in the factorization in \eqref{eq:fastalg}
was implemented in a different sub-block,
and were then wrapped together in a large module that implements the
complete transform.
Each sub-block that involves an arithmetic operation increases
the wordlength in one bit in order to avoid overflow and its outputs are registered.

\begin{figure}[!h]
\centering
\includegraphics[scale=0.8]{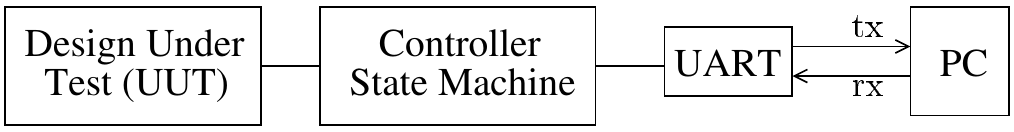}
\caption{Testbed architecture for testing the implemented designs.}
\label{fig:testbed}
\end{figure}

The designs were tested employing
the scheme shown in Fig.~\ref{fig:testbed},
together with a state-machine serving as controller and connected to a
universal asynchronous receiver-transmitter (UART) block.
The UART core interfaces with the controller state machine
using the ARM Advanced Microcontroller Bus Architecture Advanced eXtensible Interface 4
protocol.
A personal computer (PC)
communicates with the controller through the UART by
sending a packet of eight 8-bit coefficients,
corresponding to an input for the transform block.
The values of the 8-bit coefficients are
randomly generated integers in the interval~$[-10, 10]$.
The set of the eight coefficients is then passed to the design
and processed.
Then
the controller state machine
sends the eight output coefficients
back to the PC, which are then compared with the
output of a software model used to ensuring
the hardware design is correctly implemented.

Table~\ref{table:hardware}
shows the hardware resource utilization for the new transforms
in Table~\ref{T:E8}, together with the following 8-point transforms from the literature: RDCT~\cite{Cintra2011}, MRDCT~\cite{Bayer2012a}, OCBSML~\cite{Oliveira2018}, and LO~\cite{Lengwehasatit2004}.
We also compare the 8-point intDCT from HEVC~\cite{Sullivan2012}.
The displayed metrics are the
number of occupied slices,
number of look-up tables~(LUT),
flip-flop~(FF) count,
latency ($L$) in terms of clock cycles,
critical path delay~($T_{\text{cpd}}$),
maximum operating frequency $F_{\text{max}} = T_{\text{cpd}}^{-1}$,
and dynamic power~($D_p$) normalized by~$F_{\text{max}}$.

\begin{table}[!h]
\centering
	\caption{FPGA measures of the implemented architectures new and competing 8-point transforms}
	\begin{tabular}{@{ }l@{ }c@{ }c@{ }c@{ }c@{ }c@{ }c@{ }c@{ }}
	\toprule
	\multirow{3}{*}{Transform} & \multicolumn{7}{c}{Metrics} \\ \cline{2-8}
	& \multirow{2}{*}{Slices} & \multirow{2}{*}{LUT} & \multirow{2}{*}{FF} & $L$ & $T_{\text{cpd}}$ & $F_{\text{max}}$& $D_p$ \\
	& & & & (cycles) & ($\nano\second$) & ($\mega\hertz$) & ($\mu\watt\per\mega\hertz$) \\\midrule
	MRDCT~\cite{Bayer2012a} & 76 & \textbf{165} & \textbf{299} & 4 &  4.244 & 236.627 & 25.464\\
	OCBT~\cite{Oliveira2013} & 77 & 168 & 322 & 4 & \textbf{3.773} & \textbf{265.041} & 22.638 \\
	$\mathbf{{T}}_{8}^{(3)}$ & \textbf{72} & 177 & 328 & 4 & 3.991 & 250.564 & \textbf{19.995} \\
	$\mathbf{{T}}_{8}^{(4)}$ & 82 & 187 & 328 & 4 & 4.509 & 221.779 & 22.545 \\
	$\mathbf{{T}}_{8}^{(5)}$ & 96 & 234 & 421 & 5 & 4.473 & 223.564 & 26.838 \\
	RDCT~\cite{Cintra2011} & 96 & 233 & 421 & 5 & 4.043 & 247.341 & 24.258 \\
	$\mathbf{{T}}_{8}^{(7)}$ & 106 & 245 & 421 & 5 & 4.092 & 244.379 & 24.552 \\
	LO~\cite{Lengwehasatit2004} & 110 & 314 & 476  & 6 & 4.176 & 239.464 & 37.584 \\
	OCBSML~\cite{Oliveira2018} & 98 & 253 & 421 & 5 & 4.492 & 222.618 & 40.428 \\
	IntDCT (HEVC)~\cite{Sullivan2012} & 291 & 874 & 377 & \textbf{3} & 7.068 & 141.483 & 127.224 \\  \bottomrule
	\end{tabular}{\label{table:hardware}}
\end{table}

Among all considered transforms,
the proposed low-complexity matrices~$\mathbf{{T}}_{8}^{(3)}$ and~$\mathbf{{T}}_{8}^{(4)}$
display the best power efficiency.
The~$\mathbf{{T}}_{8}^{(3)}$ is the one displaying the best normalized dynamic power,
demanding about~$11.3$\% less power than the second best transform~$\mathbf{{T}}_{8}^{(4)}$,
about~$11.6$\% less than the OCBT, and $46.8$\% and~$50.54$\%
less than LO~\cite{Lengwehasatit2004} and OCBSML~\cite{Oliveira2018}, respectively.
The transform~$\mathbf{{T}}_{8}^{(3)}$ is also the one demanding the lowest number of slices,
while the MRDCT~\cite{Bayer2012a} requires the lowest number LUTs and FFs.
The transform OCBT
achieves the highest maximum operating frequency, which is followed by the new transform~$\mathbf{{T}}_{8}^{(3)}$
and the RDCT~\cite{Cintra2011}.
One can notice that LO~\cite{Lengwehasatit2004}
has one of the highest need for resources and requires the largest latency,
being outperformed in terms of speed by
the new transforms~$\mathbf{{T}}_{8}^{(3)}$ and~$\mathbf{{T}}_{8}^{(7)}$.
The OCBSML~\cite{Oliveira2018} transform is
the most inefficient in terms of normalized dynamic power,
and it is followed by LO~\cite{Lengwehasatit2004},
both requiring approximately double of the normalized power required by~$\mathbf{{T}}_{8}^{(3)}$.
IntDCT demands four times more slices and more than five times more LUTs than the best performing in these metrics ($\mathbf{{T}}_{8}^{(3)}$ and MRDCT~\cite{Bayer2012a}, respectively).
The HEVC core transform transform also has rougly doubled latency with respect to OCBSML~\cite{Oliveira2018} and demands more than six times dynamic power consumption compared to  and $\mathbf{{T}}_{8}^{(3)}$.

\section{Conclusion} \label{sec:conclusions}

This paper introduced a multiparametric class of transforms that encompasses several methods archived in literature,
such as
\cite{Cintra2011,Bayer2012a,Oliveira2013,Brahimi2020}.
The proposed formalism expands the element set that was used for proposing both the RDCT and MRDCT, by allowing transforms that
require bit-shifting operations.
We present a fast algorithm and the associated arithmetic complexity
analysis for the entire class of transforms.
To derive optimal approximations,
we set up a multicriteria optimization problem that minimizes the arithmetic complexity and the proximity to the exact DCT and maximizes the transform decorrelation capabilities.
To the best of our knowledge, this paper introduces four 8-point DCT-like transforms.
The proposed methods were comprehensively assessed and realized in hardware using FPGA technology.
We also scaled the introduced optimal 8-point DCT approximations
and obtained five 16-point and six 32-point transforms
suitable
for image and video coding.
The proposed 8-, 16-, and 32-point transforms were submitted to still-image and video compression experiments, proving to be competitive or better than state-of-the-art methods found in literature.

\section*{Acknowledgments}
We
thank for the
financial support from
Funda\c{c}\~ao de Amparo \`a Pesquisa do Estado do Rio Grande do Sul (FAPERGS),
Conselho Nacional de Desenvolvimento Cient\'{\i}fico and Tecnol\'ogico (CNPq),
and Coordena\c{c}\~ao de Aperfei\c{c}oamento de Pessoal de N\'ivel Superior (CAPES),
Brazil.

{\small
\singlespacing
\bibliographystyle{siam}
\bibliography{references}
}

\end{document}